\documentclass[aps,amsmath,amssymb,nofootinbib]{revtex4}
\usepackage{graphicx,color}
 \graphicspath{{./}{./figures/}}
\usepackage{dcolumn}% Align table columns on decimal point
\usepackage{bm}% bold math
\usepackage{braket}
\usepackage[final]{hyperref} % adds hyper links inside the generated pdf file
\hypersetup{
	colorlinks=true,       % false: boxed links; true: colored links
	linkcolor=[rgb]{0.55,0,0.5},        % color of internal links
	citecolor=[rgb]{0,0,0.6},        % color of links to bibliography
	filecolor=magenta,     % color of file links
	urlcolor=blue         
}

\definecolor{labelkey}{cmyk}{.4,.2,0,0}

\newcommand{\be}{\begin{equation}}
\newcommand{\ee}{\end{equation}}
\newcommand{\bea}{\begin{eqnarray}}
\newcommand{\eea}{\end{eqnarray}}

\newcommand{\nn}{\nonumber }

\newcommand{\JN}{{\mathbb N}}

\newcommand{\JZ}{{\mathbb Z}}
\newcommand{\JR}{{\mathbb R}}

\newcommand{\ssp}{\hspace{3pt}}

\newcommand{\spp}{{\sf p}}
\newcommand{\sZ}{{\sf Z}}
\newcommand{\sx}{{\sf x}}
\newcommand{\st}{{\sf t}}
\newcommand{\sss}{{\sf s}}

\begin{document}

\title{Midpoint distribution of directed polymers in the stationary regime:\\ exact result through linear response}

\author{Christian Maes}
\affiliation{Instituut voor Theoretische Fysica, KU Leuven}
\author{Thimoth\'ee Thiery}
\affiliation{Instituut voor Theoretische Fysica, KU Leuven}
\begin{abstract}
We obtain an exact result for the midpoint probability distribution function (pdf) of the stationary continuum directed polymer, when averaged over the disorder. It is obtained by relating that pdf to the linear response of the stochastic Burgers field to some perturbation. From the symmetries of the stochastic Burgers equation we derive a fluctuation--dissipation relation so that the pdf gets given by the stationary two space-time points correlation function of the Burgers field. An analytical expression for the latter was obtained by Imamura and Sasamoto [2013], thereby rendering our result explicit. In the large length limit that implies that the pdf is nothing but the scaling function $f_{{\rm KPZ}}(y)$ introduced by Pr\"ahofer and Spohn [2004]. Using the KPZ-universality paradigm, we find that this function can therefore also be interpreted as the pdf of the position $y$ of the maximum of the Airy process minus a parabola and a two-sided Brownian motion. We provide a direct numerical test of the result through simulations of the Log-Gamma polymer.
\end{abstract}

%\date{\today}

\maketitle

\section{Introduction}

The directed polymer (DP) problem, i.e., the equilibrium statistical mechanics of directed paths in a random environment, has over the years attracted  a considerable amount of attention from both the physics \cite{HuseHenley1985,KardarZhang1987,Kardar1987,BouchaudOrland1990} and the mathematics \cite{ImbrieSpencer1988,Bolthausen1989} community; see \cite{ThieryPHD,comets2004probabilistic,Comets2017SaintFlour} for reviews. The trade--off for the DP between disorder and interaction (elasticity) makes it a paradigmatic example for disordered systems with applications and connections to a variety of fields, ranging from vortex physics in superconductors \cite{BlatterFeigelmanGeshkenbeinLarkinVinokur1994} and domain walls in disordered magnets \cite{LemerleFerreChappertMathetGiamarchiLeDoussal1998} to population dynamics \cite{GueudreDobrinevskiBouchaud2014}. The DP exhibits an interesting low-temperature disordered phase that is reminiscent of the glassy phases observed in (even) more complex disordered systems such as spin glasses \cite{DerridaSpohn1988,FisherHuse1985}, and understanding its properties is the main challenge of the field. The DP has however also applications beyond the field of disordered systems as it is included in the Kardar-Parisi-Zhang (KPZ) universality class (KPZUC), encompassing models in different areas of out-of-equilibrium statistical mechanics such as interface growth \cite{KPZ,BarabisiStanleyBook,HalpinHealyZhang1995}, interacting particle systems \cite{KriecherbauerKrug2008}, random walks in dynamic random environment \cite{BarraquandCorwin2015,ThieryLeDoussal2016b} and many others.

The $(1+1)$D case has always been standing aside in both the DP and the KPZ context. There, the knowledge of the stationary measure of the model quickly led to the determination of the exact {\it scaling exponents} of the universality class \cite{HuseHenleyFisher1985,KardarZhang1987}. More recently, progress in the understanding of the KPZUC beyond scaling have been possible from the analysis of models with exact solvability properties, such as the Robinson-Schensted-Knuth correspondence \cite{johansson2000}, the Bethe Ansatz \cite{Kardar1987} and Macdonald processes \cite{BorodinCorwinMacDo2014}; see \cite{Corwin2011Review,QuastelSpohn2015,SpohnLesHouches2016,ThieryPHD} for reviews. The asymptotic analysis of such exact solutions reveals the existence of a remarkable universality beyond the critical exponents emerging at large scale, with the appearance of {\it universal distributions and processes} related to random matrix theory. The universal distributions were in fact found to retain some details of the boundary conditions (in the DP language) or of the initial condition (in the KPZ language), splitting the KPZUC in a number of {\it sub-universality classes}.  The three main representatives are: (i) the curved or droplet case,  where both ends of the DP are fixed; (ii) the flat case,  where one end of the DP is free to move on a line; (iii) the stationary case, corresponding for the continuum DP to the situation where one end is free to move on a line with a double-sided Brownian potential\footnote{There are also some crossover classes, see \cite{Corwin2011Review,LeDoussal2017} for review.}. Further work led to a unifying picture of these large--scale universal properties, sometimes referred to as the KPZ fixed point \cite{CorwinQuastelRemenik2011,MatetskiQuastelRemenik2016}, a picture in which the so-called {\it Airy process} \cite{PraehoferSpohn2001,ProlhacSpohn2011,QuastelRemenik2014} now plays a central role. That can for example be recognized from the results obtained for the asymptotic fluctuations of the {\it free energy of the DP} in the different sub-universality classes. In the droplet case, these were found to be distributed according to the Tracy-Widom GUE distribution  \cite{TracyWidom1993,johansson2000,CalabreseLeDoussalRosso2010,AmirCorwinQuastel2010,Dotsenko2010}, that is the one--point distribution of the Airy process. % (which is a stationary process). 
In the flat case the fluctuations of the free energy follow the Tracy-Widom GOE distribution \cite{TracyWidom1996,BaikRains2001,LeDoussalCalabrese2012,OrtmannQuastelRemenik2014}, which is in fact the distribution of the {\it maximum of the Airy process minus a parabola}. Finally, in the stationary case, one obtains the Baik-Rains distribution \cite{BaikRains2000,ImamuraSasamoto2012,BorodinCorwinFerrariVeto2015} which is the pdf of the {\it maximum of the Airy process minus a parabola and a double-sided Brownian motion}. We refer the reader to \cite{QuastelRemenik2014} for a review of these connections.

Other universal quantities of great interest in the DP-context are those directly associated with the geometric fluctuations of the DP-paths; see \cite{Halpin1991} for a numerical study. The scaling of these fluctuations with the length of the polymer has been known for a long time \cite{HuseHenleyFisher1985,KardarZhang1987}, and is quantified through the value of the roughness exponent of the DP, $\zeta = 2/3$ (superdiffusive behavior); see \cite{AgoristasLecomte2016} for a recent study. %Refined information beyond scaling similarly however had to wait for the exact--solvability revolution. 
The full pdf of the fluctuations of the {\it endpoint of the directed polymer with one end fixed and one end free to move on a line} (flat case) was  first obtained in \cite{Schehr2012,FloresQuastelRemenik2013,BaikLiechtySchehr2012} for polymers of infinite length (the universal asymptotic limit). There the connection with the Airy process appears again: the pdf is now the pdf of the {\it position of the maximum of the Airy process minus a parabola}. In the case of the $1+1$-dimensional continuum DP, the model that is studied in the present paper, the same endpoint distribution was also obtained in \cite{Dotsenko2013} through a replica Bethe Ansatz calculation for polymers of arbitrary length. Let us mention here that small--length properties of the continuum DP are also interesting from the point-of-view of universality because of the special role that it plays in the universality class: the continuum DP is the {\it universal weak noise limit} of DP--models on the square lattice \cite{AlbertsKhaninQuastel2012,BustingorryLeDoussalRosso2010}. That means that small--scale properties of the continuum DP are related to those of arbitrary DPs in a weakly disordered environment.

\medskip

In this paper we consider the stationary continuum directed polymer and obtain an exact result for the {\it distribution of the midpoint of the continuum directed polymer with stationary initial condition}, when the DP is conditioned to pass through a given point, for DPs of arbitrary lengths. Due to the special choice of the stationary initial condition, our results can also be reinterpreted as results for the {\it midpoint pdf of the DP with both ends fixed in the stationary regime}. Making the connection with KPZ universality, we show that our results lead to a prediction for the pdf of the {\it position of the maximum of the Airy process minus a parabola and a Brownian motion}. Surprisingly (to us), the obtained pdf turns out to be the well-known $f_{{\rm KPZ}}(y)$ scaling function introduced in \cite{PrahoferSpohn2004}, in agreement with a recent result obtained by Le Doussal \cite{LeDoussalToAppear2017}. We numerically check our result using simulations of the Log-Gamma polymer \cite{Seppalainen2009}, an exactly solvable DP-model on the square lattice. That provides an explicit confirmation of our result and a test of KPZ universality.

\medskip

The main point of the present paper is to find through a series of arguments that the midpoint distribution of the continuum DP with stationary initial condition is in fact equal to an {\it a priori} unrelated quantity that has already been computed using the replica Bethe Ansatz.  
As announced in the abstract, we proceed in two steps. First we relate the endpoint distribution to the linear response to some perturbation of the average field for the stationary stochastic Burgers equation. %The perturbation is some external potential.
Secondly, we show that this linear response function is equal to the two space-time points correlation function of the stochastic Burgers field in the stationary regime, i.e., we obtain a fluctuation--dissipation relation\footnote{We refer the reader to Section~\ref{Sec:Symmetries:Intro} for the relation between that part of our work and the literature.} (FDR). That correlation function is exactly known from the replica Bethe Ansatz calculation of Imamura and Sasamoto in \cite{ImamuraSasamoto2013}, which makes our result explicit. The route we follow should be contrasted with the recent work of Le Doussal \cite{LeDoussalToAppear2017} which, among other results, confirm ours for the position of the maximum of the Airy process minus a Brownian motion. Indeed, the results of \cite{LeDoussalToAppear2017} are fully based on replica Bethe ansatz calculations which a priori seem disconnected from our considerations on linear response and FDR.

The outline of this paper is as follows. In Section~\ref{Sec:MainResult} we define the models and observables we study and state our main results. Section~\ref{Sec:Symmetries} focuses on the derivation of a FDR in the stochastic Burgers equation with stationary initial condition, that is obtained through a detailed and physical discussion of the symmetries of the Martin-Siggia-Rose (MSR) action associated with the stochastic Burgers equation (see \cite{MSR,Janssen1976} for historical references on the MSR formalism and \cite{Janssen1992,TauberBook2014} for more recent reviews). Section~\ref{Sec:ApplicationToDP} contains the details of the derivation of the results concerning more particularly the midpoint distribution of the DP. We also recall in Appendix~A the exact result of Imamura-Sasamoto \cite{ImamuraSasamoto2013}, making this paper self-contained.

\section{Setting and overview of results} \label{Sec:MainResult}

\subsection{Model and units} \label{Sec:MainResult:ModelAndUnits}

\begin{figure}
\centerline{\includegraphics[width=6cm]{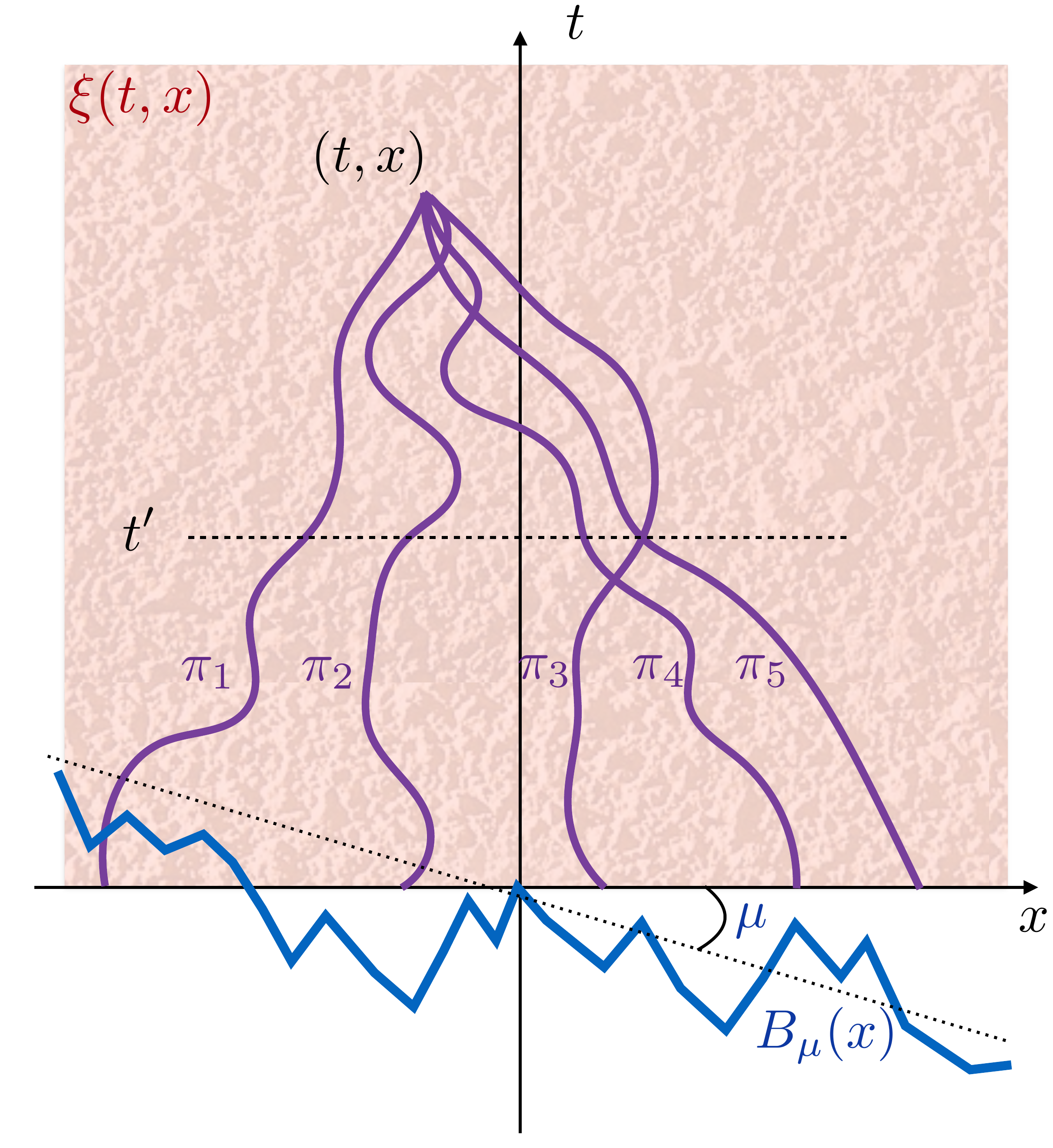}   \includegraphics[width=6cm]{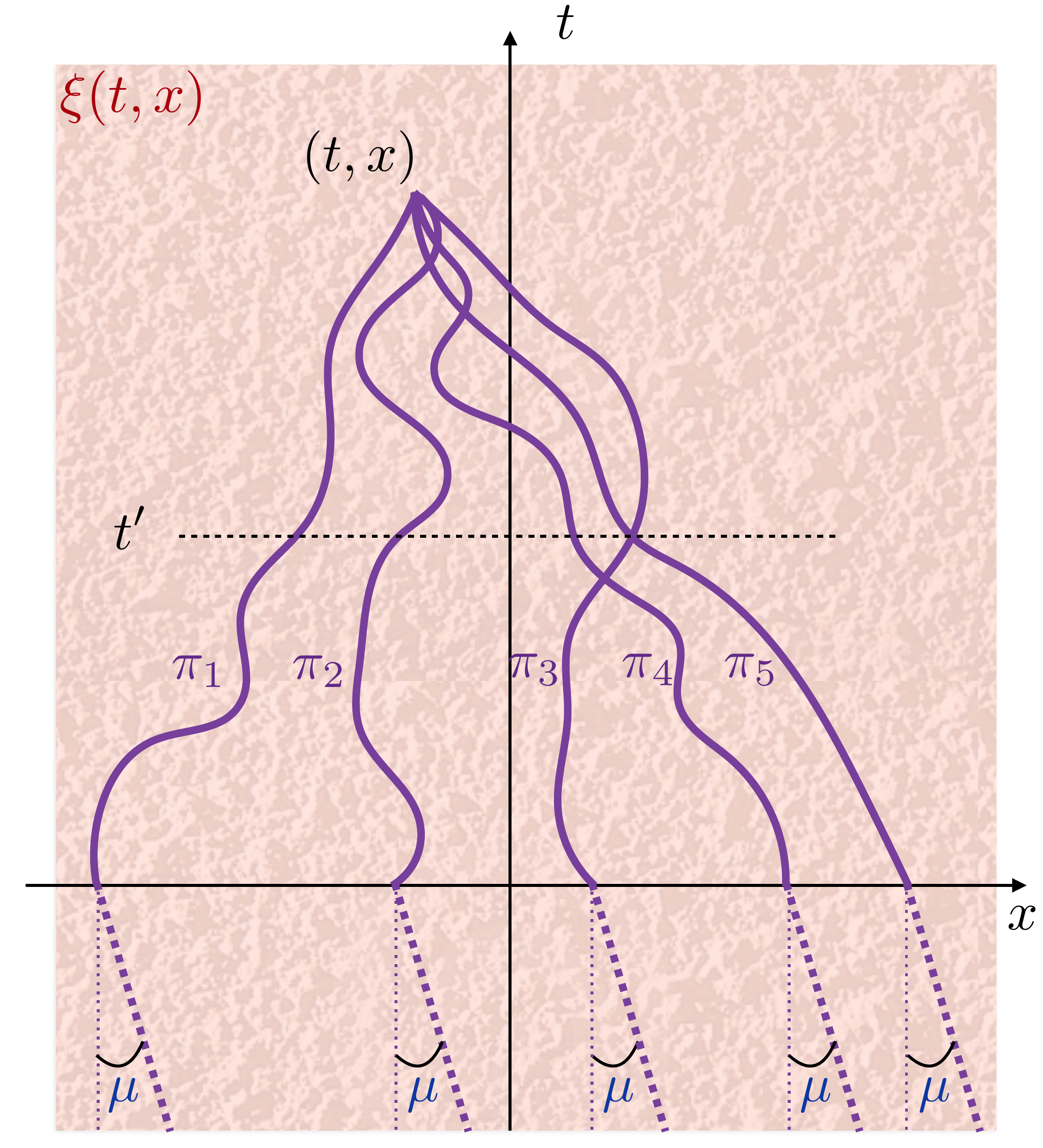}} 
\caption{Left: The continuum directed polymer with stationary initial condition at $t=0$ and endpoint at $(t,x)$. The purple lines show some admissible paths $\pi : [0,t] \to \JR$ with $\pi(0) \in \JR$ and $\pi(t) = x$. The polymer feels two types of disorder: a double-sided Brownian motion $B_{\mu}(x)$ with drift $\mu$ on the (blue) line at $t=0$ and a bulk disordered potential that is a Gaussian white noise $\xi(t,x)$ (red). The main purpose of this paper is to analyze the pdf of $x' = \pi(t')$, i.e., the point of intersection between the polymer paths (purple) and the horizontal black-dashed line. Right: Equivalently, our results also refer to the continuum directed polymer with endpoint fixed as $(t,x)$ and starting point taken as $(t_i , \mu t_i)$ with $t_i = - \infty$.}
\label{fig:DP1}
\end{figure}

We consider the continuum directed polymer under stationary conditions in $1+1$ dimension. As is conventional, we denote by $t \in \JR_+$ the longitudinal coordinate of the polymer and by $x \in \JR $ the transversal coordinate. The endpoint of the polymer is fixed at $(t,x) \in \JR_+ \times \JR$, while the initial point at $t=0$ can be anywhere on the real line. Admissible polymer paths can thus be parametrized by functions $\pi :  s \in [0,t] \to \pi(s) \in  \JR$ with $\pi(0) \in \JR$ arbitrary and $\pi(t) = x$ (see Fig.~\ref{fig:DP1}). The elasticity parameter is $\kappa >0 $ and $\xi(t,x)$ is the bulk disorder potential which is taken to be a centered Gaussian white noise with correlations $\langle \xi(t,x) \xi(t',x') \rangle =  \sigma^2  \delta(t-t') \delta(x-x')$ with $\sigma >0$ the strength of the noise. $\beta\geq 0$ will be the inverse temperature of the medium. We also suppose that there is an additional disorder on the line at $t=0$ in terms of  a double-sided Brownian motion $B_\mu(x)$ with drift parametrized by $\mu \in \JR $ as 
\bea
B_\mu(x) :=  \mu x +  \int_{0}^x dx' \eta(x')
\eea
 with $\eta(x)$  a centered Gaussian white noise with variance $\langle \eta(x) \eta(x') \rangle =  \kappa \sigma^2 \beta^3 \delta(x-x')$. As we will recall the choice of this variance confers to the model a stationarity property. Finally we also sometimes introduce an additional bulk potential $V(t,x)$ that, when present, will always be thought of as a small perturbation of the model with $V=0$. As an equilibrium statistical mechanics problem, the model consists in taking the paths weighted with the Boltzmann distribution: formally, the functional probability density $Q_{t,x}$ of a path $\pi$  is written as
\bea\label{one}
Q_{t,x}[\pi] := \frac{1}{Z (t,x)} e^{- \beta \left\{ B_\mu(\pi(0)) +  \int_{0}^t ds   \left( \frac{\kappa}{2} (\partial_s \pi(s))^2  + \xi(s,\pi(s)) + V(s, \pi(s))  \right)   \right\} } \ssp 
\eea
with partition sum given by the path-integral\footnote{This definition of the model is formal and contains some well-known caveats associated with the use of a rough disordered potential. We refer to \cite{Corwin2011Review} for more details on these issues. A more physicist-oriented discussion can also be found in \cite{ThieryPHD}.},
\bea \label{Eq:DefZPI}
Z(t,x) := \int_{\pi(0) \in \JR}^{\pi(t) = x } {\cal D}[\pi] e^{- \beta \left\{ B_\mu(\pi(0)) +  \int_{0}^t ds \left( \frac{\kappa}{2} (\partial_s \pi(s))^2  + \xi(s,\pi(s)) + V(s, \pi(s))  \right)   \right\} }  \ssp .
\eea

\medskip

{\bf Rescaling} The rescalings $t \to  \frac{1}{\kappa \sigma^4 \beta^5}\,t$, $x \to \frac{1}{\kappa \sigma^2\beta^3} \,x$, $\mu \to \kappa \sigma^2\beta^3  \,\mu $  and $V \to \beta^5\kappa \sigma^4\, V$ permit to absorb the parameters $\beta$, $\sigma$ and $\kappa$ everywhere in the above definitions. In the following we thus assume without loss of generality that the parameters of the model are  $\kappa = \sigma = \beta =1$. Hence, the only free parameter is  $\mu \in \JR$ that, as we will recall, parametrizes a family of stationary measures for the associated stochastic Burgers equation. 

\medskip

{\bf Notation for averages} Throughout the rest of the paper we use the notation $\langle \cdot \rangle_{\mu}^{V}$ to denote the average over both $\eta$ and $\xi$ for a given choice of $\mu$ and $V$. In particular $\langle \cdot \rangle_{\mu}^{0}$ refers to the average in the absence of the perturbing potential. When no confusions is possible, we put  $\langle \cdot \rangle$ for the average over any disorder that is present.

\subsection{Some associated stochastic evolution equations} \label{Sec:MainResult:SPDES}

It is well-known \cite{Corwin2011Review,ThieryPHD,SpohnLesHouches2016,QuastelSpohn2015} that the equilibrium statistical mechanics model of the DP above is associated with some stochastic partial differential equations where $t$ plays the role of time. In particular the partition sum $Z (t,x) $ satisfies the multiplicative-stochastic-heat-equation
\bea \label{Eq:Main:MSHE}
&& \partial_t Z(t,x) = \frac{1}{2} \partial_x^2 Z(t,x) -( \xi(t,x) + V(t,x)) Z(t,x) \nn \\
&& Z(0,x) = e^{-B_\mu(x)}  \ssp .
\eea
to be interpreted in the It\^o--sense (see \cite{Corwin2011Review,ThieryPHD} for a discussion of this important subtlety), as will be the case for all the stochastic equations considered in this paper. The logarithm $h(t,x) := \ln Z(t,x)$ satisfies the  Kardar-Parisi-Zhang (KPZ) equation (again, see \cite{Corwin2011Review,ThieryPHD}):
\bea \label{Eq:Main:KPZ}
&& \partial_t h(t,x) =  \frac{1}{2} (\partial_x h(t,x))^2  + \frac{1}{2} \partial_x^2 h (t,x) -  \xi(t,x) - V(t,x)  \nn \\
&& h(0,x) = -B_\mu(x) \ssp .
\eea
That equation was originally introduced in \cite{KPZ} to describe the out-of-equilibrium stochastic growth of an interface and throughout the paper we will refer to $h$ as the height of the interface which is thus also minus the free energy of the directed polymer. Finally the slope field $u(t,x) := \partial_x h(t,x)$ satisfies the stochastic Burgers equation: with $\phi(t,x) := - \partial_x V(t,x)$,
\bea \label{Eq:Main:Burgers}
&& \partial_t u(t,x) =  \frac{1}{2} \partial_x ( u(t,x))^2  + \frac{1}{2} \partial_x^2 u (t,x) -  \partial_x \xi(t,x) + \phi(t,x)  \nn \\
&& u(0,x) = -\mu - \eta(x) \ssp .
\eea
Recall that the stochastic Burgers equation (\ref{Eq:Main:Burgers}) also governs the time-evolution of the density of particles in the weakly asymmetric exclusion process (WASEP). Then, $\rho(t,x):=1/2+u(t,x)$ is a local density of particles, and (\ref{Eq:Main:Burgers}) is the continuum limit of a lattice gas on $\JZ$ where particles perform a biased random walk and interact through a hard-core repulsive potential. The presence of a bias in that discrete diffusion manifests itself in (\ref{Eq:Main:Burgers}) through the presence of the nonlinear term $ \frac{1}{2} (\partial_x u(t,x))^2 $.

\medskip

The important property of the chosen initial condition is that in the absence of the perturbation $\phi(t,x)$, the process $u(t,x)$ is stationary; see \cite{ForsterNelsonStephen1977,HuseHenleyFisher1985}:  $u(t,x) \sim u(0,0)$ in law. That will also follow in Section~\ref{Sec:Symmetries}. Note that in any case that does not mean that $h$ or $Z$ in \eqref{Eq:Main:MSHE}--\eqref{Eq:Main:KPZ} are stationary; in fact the interface grows and $\lim_{t \to \infty } h(t,x) = + \infty$ (almost surely). Still, $h(t,x) - h(t,0) = \ln (Z(t,x)/Z(t,0)) $ is stationary and is given in law by $- B_\mu(x)$.

\subsection{Linear response in Burgers and in the DP problem through a FDR} \label{Sec:MainResult:LinRes}

\begin{figure}
\centerline{\hspace{2cm} \includegraphics[width=9cm]{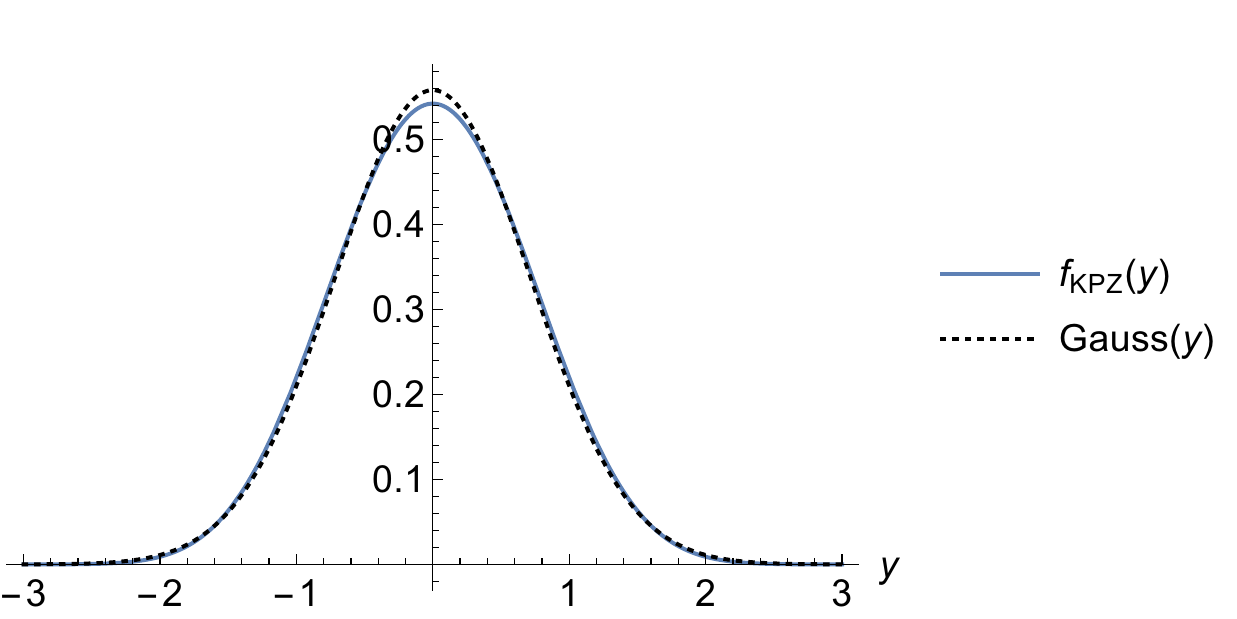}} 
\caption{The $f_{\rm KPZ}(y)$ scaling function, compared with a Gaussian with the same second moment. The plot of $f_{\rm KPZ}(y)$ was realized using the table available in \cite{PrahoferFunctions}}.
\label{fig:fKPZ}
\end{figure}

In Section~\ref{Sec:Symmetries} we consider the following linear response function for the perturbation of the average value of the slope field $u(t,x)$ by a small source $\phi(t',x')$ as in \eqref{Eq:Main:Burgers}:
\bea \label{Eq:Main:DefR}
R_{\mu}((t,x);(t',x')) :=  \frac{\delta \langle  u(t,x) \rangle_{\mu}^\phi }{ \delta \phi(t',x')} |_{\phi=0}  \ssp .
\eea
Studying the symmetries of the MSR-action of the theory, we show that this response function is stationary: $R_{\mu}((t,x);(t',x')) \equiv R_{\mu}(t-t',x-x')$, satisfies $R_{\mu}(t-t',x-x') = R_{0}(t-t',x-x' - \mu(t-t'))$ and is given by the {\it fluctuation--dissipation relation},
\bea \label{Eq:MainResultFDT}
R_{0} (t-t',x-x') =  \theta(t-t')\, \langle u(t,x) u(t',x') \rangle_{0}^0 \ssp .
\eea
We note that this relation was already stated in \cite{ForsterNelsonStephen1977} (based on \cite{DekerHaake1975}), but a complete physical discussion of this relation and of its derivation seemed to be lacking in the KPZ context (such relations are usually the hallmark of systems at thermodynamic equilibrium, while the KPZ equation definitely represents an out-of-equilibrium situation, see Sec.~\ref{Sec:Symmetries:Intro}).
The right-hand side of (\ref{Eq:MainResultFDT}), i.e., the Burgers stationary two-point correlation function, is known explicitly from the work of Imamura and Sasamoto in \cite{ImamuraSasamoto2013}. The latter is recalled for completeness in Appendix~\ref{app:IS}.  Making contact with their notation we show in particular in Appendix~\ref{app:IS} that rescaling the response function as
\bea \label{Eq:MainResultRRescaling}
{\cal R}_0(t,y) := (2 t^2)^{1/3} R_0(t, (2 t^2)^{1/3} y  ) \ssp ,
\eea
gives
\bea \label{Eq:MainResultFDTRescaled}
{\cal R}_0(t,y) =  \frac{1}{4} g_t''(y) \ssp ,
\eea
with $g_t(y)$ the function whose expression was given in Corollary~4 of \cite{ImamuraSasamoto2013}, and which is recalled in Appendix~\ref{app:IS}. In the large--time scaling limit Imamura and Sasamoto argued in \cite{ImamuraSasamoto2013} that $\frac{1}{4} g_t''(y)$ converges as $t \to \infty$ to the scaling function $f_{{\rm KPZ}}(y)$ introduced by Pr\"ahofer and Spohn in \cite{PrahoferSpohn2004} (there simply denoted as $f(y)$) and widely encountered in the KPZ universality class. In our interpretation this thus means that the response function \eqref{Eq:MainResultFDT}--\eqref{Eq:Main:DefR} rescales asymptotically as
\bea\label{fy}
\lim_{t \to \infty}\,(2 t^2)^{1/3} R_0(t, (2 t^2)^{1/3} y  )   = f_{{\rm KPZ}}(y) \ssp .
\eea
A plot of $f_{{\rm KPZ}}(y)$ is given in Fig.~\ref{fig:fKPZ}. We  recall some of its known properties in Section~\ref{Sec:MainResult:Midpoint2}.

\medskip

In Section~\ref{Sec:ApplicationToDP} we show that the response function $R_\mu$ is also the linear response of the free energy of the DP (KPZ-height field) to a small external potential:
\bea \label{Eq:MainLinResDP}
R_{\mu} (t-t',x-x') = - \frac{\delta \langle  \ln Z(t,x) \rangle_{\mu}^V }{ \delta V(t',x')}|_{V=0}  \ssp .
\eea
{}
We thus obtain an exact formula for the linear response of the averaged free energy to a small external potential, a very natural question.
Yet,  in the present paper we show that the left-hand side is also related to the {\it midpoint distribution of the directed polymer in the stationary regime}. We explore the details in Section~\ref{Sec:ApplicationToDP} and the results  are summarized in the remaining of this section.

\subsection{Midpoint distribution of the DP: arbitrary times} \label{Sec:MainResult:Midpoint1}

We come back to the setting of Section~\ref{Sec:MainResult:ModelAndUnits}, i.e., of the DP with stationary initial condition at $t=0$ and a fixed endpoint at some $(t,x)$ in the absence of an external perturbation potential $V$.
\smallskip

Admissible polymer paths are thus continuous functions $\pi : [0,t] \to \JR$ with $\pi(0) \in \JR$ and $\pi(t) = x$. Introducing an intermediary time $t' \in [0,t]$, we are interested in the pdf of $x' = \pi(t')$ in the model without the external potential, i.e., when $V(t,x) = 0$. In a fixed random environment the stationary probability that the directed polymer passes through the point $(t',x')$ knowing that it passes through $(t,x)$ (see Fig.~\ref{fig:DP1}), with $0<t'<t$ is given via \eqref{one} as,
\bea \label{Eq:Defq}
q((t',x') | (t,x) )  := \int_{\pi(0) \in \JR}^{\pi(t)=x} {\cal D}[\pi]\, \delta(\pi(t')-x')\, Q_{t,x}(\pi) \ssp .
\eea
In this paper we consider its average over disorder,
\bea
&& p_{\mu}((t',x')|(t,x)):= \langle q((t',x') | (t,x) )  \rangle_\mu^{V=0}   \\
&& = \left\langle \frac{\int_{\pi(0) \in \JR}^{\pi(t) = x } {\cal D}[\pi]  \delta( \pi(t') -x') e^{-  \left\{ B_\mu(\pi(0)) +  \int_{0}^t ds \left( \frac{1}{2} (\partial_s \pi(s))^2  + \xi(s,\pi(s))   \right)   \right\} } }{Z(t,x)}  \right\rangle_{\mu}^{V=0}  \nn \ssp .
\eea
and we obtain it by relating in Section~\ref{Sec:ApplicationToDP} the pdf to the response function previously introduced, i.e., we show that
\bea
p_{\mu}((t',x')|(t,x)) = R_{\mu}(t-t',x-x') \ssp .
\eea
Hence the results previously presented in the context of the linear response problem can now be translated to yield an explicit expression for $p_\mu((t',x')|(t,x))$. In particular from \eqref{Eq:MainResultFDT} we thus obtain $p_\mu(t-t',x-x') = p_0(t-t',x-x'-\mu(t-t')) $ with
\bea \label{Eq:Mainresult}
p_{0}(t-t',x-x') = \langle u(t,x) u(t',x') \rangle_{0}^0 \ssp ,
\eea
which is our main result. Furthermore, rescaling $p_0$ as
\bea \label{Eq:MainResRescaling}
{\cal P}(t , y) :=  (2 t^2)^{1/3} p_0(t, (2 t^2)^{1/3} y  )  \ssp , 
\eea
we use \eqref{Eq:MainResultRRescaling}--\eqref{Eq:MainResultFDTRescaled} to get
\bea \label{Eq:MainresultRescaled}
{\cal P}(t , y) = \frac{1}{4} g_t''(y) \ssp ,
\eea
%where the meaning of $g_t(y)$ was recalled in Section~\ref{Sec:MainResult:LinRes}. 
Note that the rescaling (\ref{Eq:MainResRescaling}) is indeed natural since we retrieve here the DP roughness exponent $\zeta =2/3$, expressing that paths fluctuate on the scale $\langle \pi((t-\tau))^2  \rangle_0^0 - (\langle \pi((t-\tau))  \rangle_0^0)^2 \sim \tau^{2 \zeta} $. 

\subsection{Midpoint distribution of the DP: large--time limit and \\the solution of a variational problem involving the Airy process} \label{Sec:MainResult:Midpoint2}

 Following our above results, we thus deduce that the rescaled probability ${\cal P}(t , y) $ converges as in \eqref{fy} to
\bea \label{Eq:MainResult:ProbafKPZ}
{\cal P}_{\infty} (y) := \lim_{t \to \infty} {\cal P}(t , y)  = f_{{\rm KPZ}}(y) \ssp ,
\eea
where $f_{{\rm KPZ}}(y)$ is the scaling function introduced by Pr\"ahofer and Spohn in \cite{PrahoferSpohn2004} and which is plotted in Fig.~\ref{fig:fKPZ} --- it is symmetric and it decays at large scale as $f_{{\rm KPZ}}(y) \sim e^{-c |y|^3}$ with $c \simeq 0.295(5)$ \cite{PrahoferSpohn2004}, distinguishing it from a Gaussian distribution. That must be compared to the asymptotic decay of the (rescaled) endpoint pdf of the DP with one end fixed and one end free to move on a line: there the exact decay was found to be $\sim e^{-y^3/12}$ \cite{Schehr2012}; see also \cite{BothnerLiechty2013} for the subleading corrections. We thus find the same stretched--exponential decay for the stationary midpoint distribution but with a different numerical prefactor. %This $e^{-y^3}$ decay thus appears very universal in the DP problem.
 Using the table of $f_{{\rm KPZ}}(y)$ available in \cite{PrahoferFunctions} we also estimate $f_{{\rm KPZ}}(0) \simeq 0.54$ and the second and fourth moment of the $f_{{\rm KPZ}}(y)$ scaling function as $\sigma_{{\rm KPZ}}  = \sqrt{ \int_{y \in \JR } y^2 f_{{\rm KPZ}}(y) dy}  \simeq 0.714 $ and $\langle y^4 \rangle = \int_{y \in \JR } y^4 f_{{\rm KPZ}}(y) dy \simeq 0.733 $. The kurtosis of the distribution is approximately $\frac{\langle y^4 \rangle}{ \sigma_{{\rm KPZ}}^4} \simeq 2.812$, distinguishing it again from a simpler Gaussian where the kurtosis is exactly $3$.

\medskip

In Section~\ref{Sec:ApplicationToDP:Airy} we use the conjectural relation between the large scale statistical properties of the continuum DP and the Airy process, as is reviewed in \cite{QuastelRemenik2014}.  From \eqref{Eq:MainResult:ProbafKPZ} then follows that
$f_{{\rm KPZ}}$ can also be seen as the pdf of the point at which the maximum of a variational problem involving the Airy process ${\cal A}_2(y)$ introduced by Pr\"ahofer and Spohn in \cite{PraehoferSpohn2001} is attained.  We show that if $y$ has pdf $f_{{\rm KPZ}}(y)$, then
\bea \label{Eq:MainRes:Airy}
y \sim {\rm argmax}_{z \in \JR} \left[   {\cal A}_2(z)  - z^2   -  \sqrt{2} B(z)) \right]  \ssp \text{ in law}
\eea
with $B(z) = \int_{0}^z \eta(z') dz' $ a standard unit double-sided Brownian motion independent of ${\cal A}_2(z)$. %This is done by using the conjectural relation between the large scale statistical properties of the continuum DP and the Airy process, as is reviewed in \cite{QuastelRemenik2014}.

\medskip

{\bf Remark} Note that since its introduction by Pr\"ahofer and Spohn in \cite{PrahoferSpohn2004} the function $f_{{\rm KPZ}}$ was known  to be a probability distribution function. Indeed, there $f_{{\rm KPZ}}$ is obtained as the scaling limit of the pdf of the position of a second class particles in a polynuclear growth model. Still in \cite{PrahoferSpohn2004}, this pdf is argued to be equal to the two-point correlation function of the density in the TASEP, a correspondence which then allows exact calculations following \cite{PrahoferSpohn2002}. There is thus a remarkable similarity between some aspects of \cite{PrahoferSpohn2004} and our study. It would be interesting to understand if there is more to this than a beautiful coincidence.

\subsection{Stationary midpoint distribution} \label{Sec:MainResult:Midpoint3}

Up to now our results were presented in the framework of the directed polymer with a stationary initial condition. Among possible initial condition, the stationary initial condition is special since it is, in some sense, retrieved at large scales starting from any initial condition. In particular in Section~\ref{Sec:ApplicationToDP:PointToPoint} we consider the continuum directed polymer problem with both ends fixed: starting point at $(t_i,x_i)$ and endpoint at $(t,x)$.  We consider the averaged ``midpoint'' pdf $p((t',x') |(t_i,x_i) , (t,x))$ that the  DP passes at $(t',x')$, with $t_i <t' <t$:
\be \label{Eq:Main:BeyondStat00}
p((t',x') |(t_i,x_i) , (t,x)) := \left\langle \frac{ \int_{\pi(t_i) = x_i}^{\pi(t) = x}  {\cal D}[\pi] \delta(\pi(t') -x') e^{-  \int_{t_i}^t ds   \left( \frac{1}{2} (\partial_s \pi(s))^2  + \xi(s,\pi(s))   \right)   } }{\int_{\pi(t_i) = x_i}^{\pi(t) = x}  {\cal D}[\pi]  e^{-  \int_{t_i}^t ds   \left( \frac{1}{2} (\partial_s \pi(s))^2  + \xi(s,\pi(s))   \right)   }} \right\rangle \ssp ,
\ee
Here the average is only over the bulk disordered potential $\xi$. Taking the initial point as $(t_i ,x_i = \phi t_i)$ with $t_i \to - \infty$ (large scale limit), and keeping $(t,x)$ and $(t',x')$ fixed we argue in Section~\ref{Sec:ApplicationToDP:PointToPoint} that the following limit holds
\bea \label{Eq:Main:BeyondStat}
\lim_{t_i \to - \infty}  p((t',x') |(t_i, \varphi t_i ) , (t,x)) =  p_{\mu = \varphi}((t',x') | (t,x)) \ssp .
\eea
Therefore our results apply to the {\it midpoint distribution of the continuum directed polymer in the stationary regime; cf. right figure in Fig.~\ref{fig:DP1}.} 

\smallskip

{\bf Remark} From this interpretation, the fact that the midpoint distribution of the directed polymer in the stationary regime differs from the distribution of the endpoint (as emphasized in Sec.~\ref{Sec:MainResult:Midpoint2} in the large length limit) might seem surprising at first glance. The intuition is partially correct: in that case since the initial point of the polymer is infinitely far from the midpoint, the midpoint indeed does not feel (apart from an eventual drift) the elastic force induced by the pinning of the initial point. However, the midpoint does feel an effective disorder that takes into account the microscopic disorder felt by the polymer from $t_i$ to $t'$. The latter depends on $x'$ and can exactly be taken into account by adding a two sided Brownian motion potential at $t'$ as in the initial formulation of the problem. This difference with the endpoint fluctuations should be particularly clear from the characterization of the asymptotic midpoint fluctuations in terms of the Airy process, see Eq.~\eqref{Eq:MainRes:Airy}.

\subsection{Universality of the result and a numerical check in the Log-Gamma polymer} \label{Sec:Main:LG}

We now discuss the universality of the result for other DP-models. For simplicity we discuss  the Log-Gamma polymer, an exactly solvable model on the square lattice introduced in \cite{Seppalainen2009}, but the discussion can be adapted to other models. As we will discuss, the interesting aspect of considering this model is that enough is known about it \cite{Seppalainen2009,ThieryLeDoussal2014,CorwinOConnellSeppalainenZygouras2014,BorodinCorwinRemenik2013} so that we obtain a prediction that does not contain any unknown scaling parameters. As is usual in this context we will not speak about the disorder as given by some random energies $E \in \JR$, but rather directly consider the random Boltzmann weights $w = e^{-\beta E} \in \JR_+$. We consider the stationary version of the model \cite{Seppalainen2009,Thiery2016} % but results beyond this setting could also be stated as in Section~\ref{Sec:MainResult:Midpoint3}, and also 
and we restrict ourselves for simplicity to the case where the stationary measure is unbiased ($\mu=0$).

\begin{figure}
\centerline{\includegraphics[width=8cm]{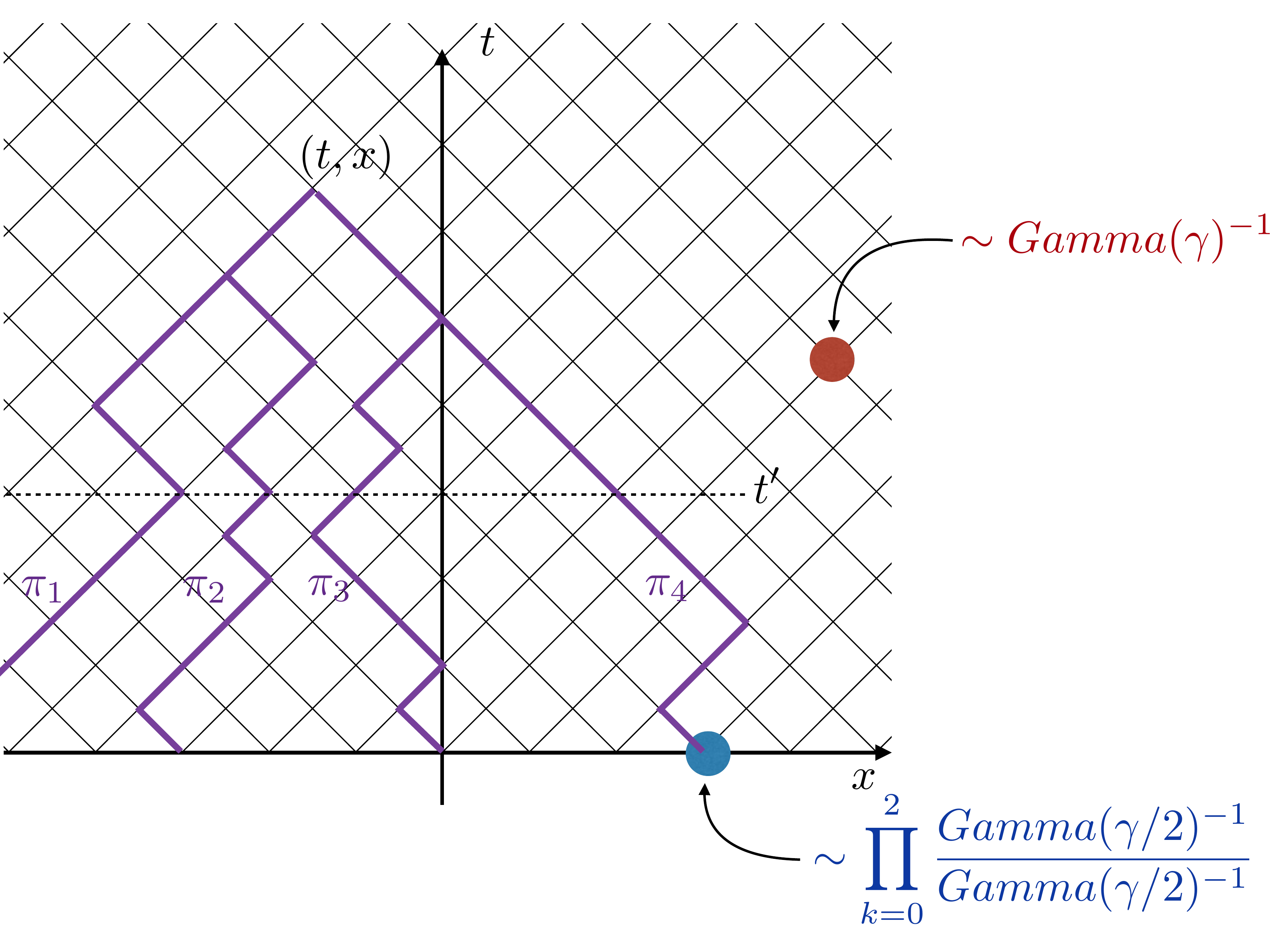}} 
\caption{The stationary Log-Gamma polymer. The model is now on the square lattice and admissible paths (purple lines) satisfy the constraint $\pi(s+1)-\pi(s) = \pm 1/2$, $\pi(0) \in \JZ$ and $\pi(t)=x$. The bulk (red) and initial (blue) disorder are taken so that the model is stationary in the same sense as it is in the continuum model. The probability distribution of the position of the intersection of the DP with the line at $t'$ coincides in two scaling limits with the equivalent quantity computed in the continuum model, see Eq.~(\ref{Eq:MainRes:WeakU}) and (\ref{Eq:MainRes:StrongU}).}
\label{fig:LG}
\end{figure}

The stationary Log-Gamma polymer can be defined as follows. Considering the square lattice with the $(\st,\sx)$ coordinates as depicted in Fig.~\ref{fig:LG}, we assign to each vertex $(\st,\sx) \in  \JN_0 \times \JZ $ in the bulk a random Boltzmann $w_\st(\sx) \in \JR_+$. These are independent, identically distributed (iid) random variables (RVs) distributed as the inverse of gamma RVs
\bea
w_\st(\sx) \sim (Gamma(\gamma))^{-1} \ssp,
\eea
where $\gamma>0$ is a parameter of the model, and we recall that a RV $w$ is distributed as $w \sim Gamma(\gamma)$ if its pdf is $p(w) = \theta(w) \frac{1}{\Gamma(\gamma)}w^{-1+\gamma}e^{-w}$ where $\Gamma$ is Euler's gamma function. On the initial time $t=0$, the disordered Boltzmann weights are distributed as
\bea
w_0(0) =1 \quad , \quad w_0(\sx \geq 1) = \prod_{k=0}^{\sx-1} \frac{u(k)}{v(k)} \quad , \quad w_0(\sx \leq -1) = \prod_{k=1}^{-\sx} \frac{v(-k)}{u(-k)} \ssp ,
\eea
where the $u(k)$ and $v(k)$ for $k \in \JZ$ are two sets of iid random variables distributed as $u \sim v$ and
\bea
u \sim v \sim (Gamma(\gamma/2))^{-1} \ssp .
\eea
Finally, the partition sum is given by the sum over paths
\bea
\sZ(\st,\sx) := \sum_{\pi, \pi(0)\in \JZ , \pi(\st)=\sx} \prod_{\sss = 0}^\st w_{\sss}(\pi(\sss)) \ssp ,
\eea
where the sum is over directed paths, i.e., paths with $\pi(\st'+1) - \pi(\st') = \pm 1/2$ (see Fig.~\ref{fig:LG}). Similarly as in the continuum DP case, here the initial condition is chosen \cite{Seppalainen2009} so that the increments of the free energy $\ln \sZ(\st,\sx+1)-\ln \sZ(\st,\sx)$ are independent for fixed $\st$, stationary and distributed as $\ln \sZ(\st,\sx+1)-\ln \sZ(\st,\sx) \sim \ln(u) - \ln(v)$ where $u$ and $v$ are two iid random variable distributed as $u\sim v \sim (Gamma(\gamma/2))^{-1}$ (the free energy of the directed polymer is thus a discrete random walk at each $\st$).

\smallskip

Similarly as in the continuum case, we consider the average over disorder of the probability that the DP passes through $(\st' , \sx')$, knowing it ends at $(\st,\sx)$

\bea
\spp_0((\st' , \sx') |(\st , \sx)) := \langle \frac{1}{\sZ(\st,\sx)} \sum_{\pi, \pi(0)\in \JZ , \pi(\st')= x' , \pi(\st)=\sx}  \ssp \prod_{\sss = 0}^\st w_{\sss}(\pi(\sss))  \rangle \ssp
\eea
where here the average is over the discrete random Boltzmann weights. As in the continuum case, it is clear from the stationarity property of the model that $\spp_0((\st' , \sx') |(\st , \sx))$ is just a function of $\st-\st'$ and $\sx-\sx'$: $\spp_0((\st' , \sx') |(\st , \sx)) \equiv \spp_0(\st-\st',x-x')$, and we can write
\bea \label{Eq:Main:DefpLG}
\spp_0(\st,\sx) =  \langle \frac{1}{\sZ(\st,\sx)} \sum_{\pi, \pi(0)=0 , \pi(\st)=\sx}  \ssp \prod_{\sss = 0}^\st w_{\sss}(\pi(\sss))  \rangle \ssp .
\eea

\smallskip

As we discuss now there are two senses in which our results on the continuum directed polymer are universal and can be applied to this discrete model: a weak and a strong universality.

\smallskip

{\bf Weak universality} The weak universality of the continuum directed polymer is the fact that it is the universal weak-noise scaling limit of discrete models of directed polymers on the square lattice. In the Log-Gamma polymer case, the weak-noise limit consists in taking the $\gamma \to \infty$ limit (small variance of the disorder) while taking a diffusive rescaling of the discrete coordinates. In Section~\ref{Sec:ApplicationToDP:LG} we argue that
\bea  \label{Eq:MainRes:WeakU}
\lim_{\gamma \to \infty} \frac{\gamma}{2} \spp_0(\st= \gamma^2 t , \sx = \frac{\gamma}{2} x ) = p_0(t,x) \, ,
\eea
with $p_0$ the midpoint distribution in the continuum model, given by Eq.~\eqref{Eq:Mainresult}. {\it Our finite time result in the continuum setting can thus be applied to a large time diffusive scaling limit in a weakly disordered environment for a discrete model.}

\smallskip

\begin{figure}
\centerline{\includegraphics[width=7cm]{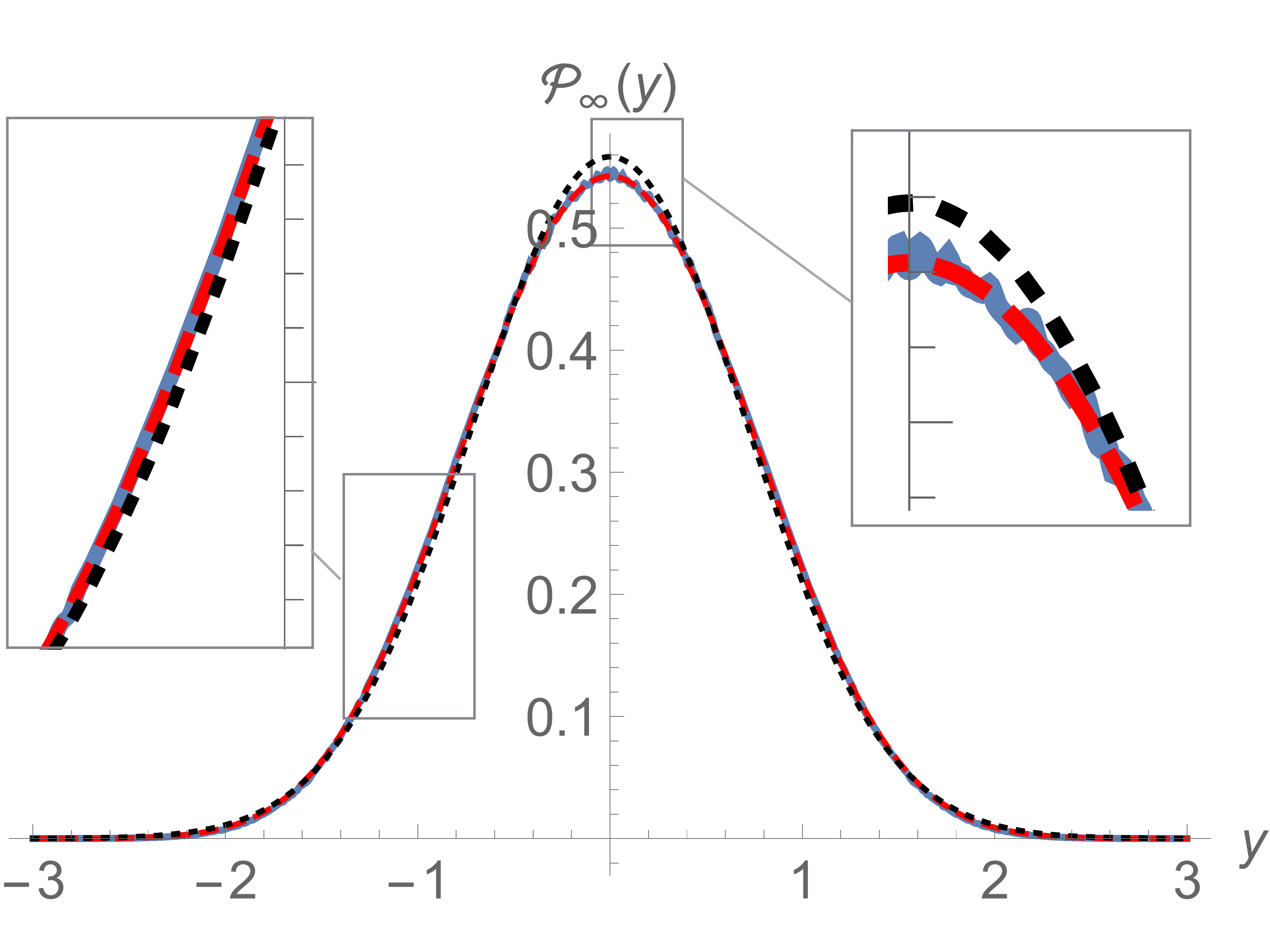}  \includegraphics[width=6.5cm]{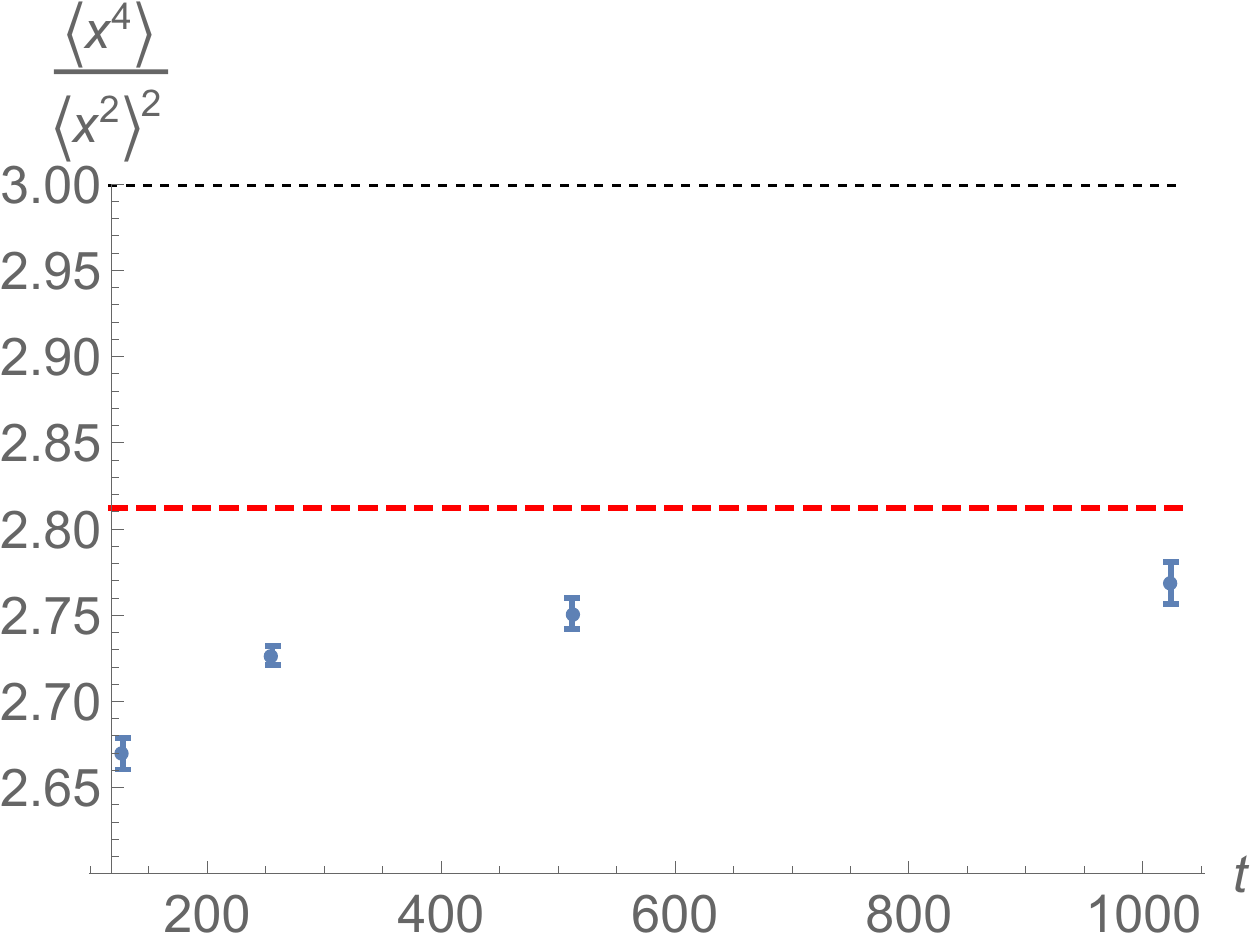}} 
\caption{Left: In blue the numerical approximation of ${\cal P}_{\infty}(y)$ using simulations of the stationary Log-Gamma polymer for polymers of length $\st = 1024$ and using the result (\ref{Eq:MainRes:StrongU}). That is compared to the red dashed line which is a plot of the $f_{{\rm KPZ}}(y)$ scaling function using the table available at \cite{PrahoferFunctions}.  The black dotted line is a Gaussian distribution with the same mean-square displacement as $f_{{\rm KPZ}}(y)$ (numerically evaluated as $\sqrt{\int_{y \in \JR} y^2 f_{{\rm KPZ}}(y) dy } \simeq 0.714$). There are no fitting parameters. Right: The blue dots give the numerical evaluation of the Kurtosis of the midpoint distribution $\spp_0(\st,\sx) $ in the Log-Gamma polymer for polymers of size $\st = 128,256,512,1024$. The red-dashed line corresponds to the numerical evaluation of the kurtosis of the $f_{{\rm KPZ}}(y)$ distribution as explained in Section~\ref{Sec:MainResult:Midpoint2}. The black dotted line corresponds to the kurtosis of a Gaussian distribution. Error-bars are $3-\sigma$ estimates.}
\label{fig:SimuMainRes}
\end{figure}

{\bf Strong universality } The strong universality conjecture is that, up to some non-universal rescaling factors, the large scale behavior of all models in the KPZ universality class, and in particular of DP-models on the square lattice, is identical. This means in particular that {\it the (properly rescaled) large $\st$ limit of $\spp_0(\st,\sx)$ should be identical to the rescaled large $t$ limit of $p_0(t,x)$}. In the scaling limit $\sx \sim \st^{2/3}$ with $\st \to \infty$, the properties of the discrete model corresponds to those of the continuum in the same scaling limit $x \sim t^{2/3}$ with $t\to \infty$. Some additional rescalings are however still necessary to set the non-universal constants (in the same way that we rescaled the parameter $\beta,\kappa$ and $\sigma$ of the continuum model in Section~\ref{Sec:MainResult:ModelAndUnits}). Writing $\sx = \kappa_{\sx} x$ and $\st = \kappa_{\st} t$ with $\kappa_{\sx} , \kappa_{\st}>0$, the two scaling parameters are chosen to ensure that: (i) the curvature of the free energy of the point to point discrete and continuum model coincide in the scaling limit; (ii) the variance of the differences of free energies between two different points in the stationary models coincide\footnote{The (i) and (ii) requirements can be respectively thought of as setting the parameter in front of the parabola and choosing the variance of the Brownian motion and Airy process to be as in Eq.~(\ref{Eq:MainRes:Airy})}. In Section~\ref{Sec:ApplicationToDP:LG} we argue that this can be done by taking 
\bea \label{Eq:Main:RescalingLogGamma2}
\kappa_{\sx} = \frac{1}{2 \psi'(\gamma/2)}   \quad , \quad \kappa_{\st} = - \frac{1}{\psi''(\gamma/2)} \ssp ,
\eea
where $\psi(x) = \frac{\Gamma'(x)}{\Gamma(x)}$ is the diGamma function. This then leads to the conjecture that, using the scaling limit already used in the continuum model, see (\ref{Eq:MainResRescaling}) and (\ref{Eq:MainResult:ProbafKPZ}),
\bea \label{Eq:MainRes:StrongU}
\lim_{\st \to \infty} (2 (\st/\kappa_\st)^2)^{1/3}  \kappa_\sx  ~ \spp_0(\st , (2 (\st/\kappa_\st)^2)^{1/3}) \kappa_{\sx}  y)   = {\cal P}_\infty(y) = f_{{\rm KPZ}}(y) \ssp .
\eea
That can be easily checked using direct numerical simulations of the stationary Log-Gamma polymer, see Fig.~\ref{fig:SimuMainRes} where we obtain an excellent agreement. We refer the reader to Section~\ref{Sec:ApplicationToDP:LG} for more details on the simulations and some complementary numerical results.

\medskip

{\bf Remark} We discussed here the universality of our results in the context of the exactly solvable Log-Gamma polymer. The weak universality result (\ref{Eq:MainRes:WeakU}) could in fact be stated in full generality since it does not rely on the exact solvability of the model. On the other hand, for the strong universality result (\ref{Eq:MainRes:StrongU}) we took advantage of the exact solvability of the model to obtain the constants $\kappa_{\st}$ and $\kappa_{\sx}$ (\ref{Eq:Main:RescalingLogGamma2}) analytically, therefore permitting a comparison with numerics without the need of fitting any parameter. Similar (and as precise) results, could be obtained in models where both the limiting free energy and the stationary measure are known, as is the case for the Strict-Weak \cite{CorwinSeappalainenShen2015} and Inverse-Beta polymers \cite{ThieryLeDoussal2015,Thiery2016}. For other models, the results could also be stated by {\it defining} $\kappa_{\sx}$ and $\kappa_{\st}$ through an independent measurement of the limiting free energy and the stationary measure of the model. 

\section{Symmetries of the stochastic Burgers equation in the stationary regime and a fluctuation-dissipation relation} \label{Sec:Symmetries}

\subsection{Preliminary comments} \label{Sec:Symmetries:Intro}

The main goal of this section is to derive the fluctuation--dissipation relation (FDR) Eq.~(\ref{Eq:MainResultFDT}). We will thus momentarily forget about the directed polymer problem and rather focus on the stochastic Burgers equation Eq.~(\ref{Eq:Main:Burgers}). The derivation will be based on the identification of some symmetries of the MSR action (see \cite{TauberBook2014} for a review of the MSR formalism) associated to the stochastic Burgers equation. The relation between the FDR or more general response formul{\ae} and the action governing dynamical ensembles has been pioneered in \cite{BaiesiMaesWynants2009,BaiesiMaes2013,BasuKrugerLazarescuMaes2015}. Our study of the symmetries of the MSR action is also inspired by \cite{AronBiroliCugliandolo2010} where that is done in detail in the rather general setting of Langevin equations with non-conserved colored multiplicative noise (which however does not apply to our case).

Let us first review some related results that exist in the literature. We note that  time-reversal symmetries have already been discussed in the context of the KPZ equation, see e.g. \cite{FreyTaeuber1994,FreyTaeuberHwa1996,CanetChateDelamotteWschebor2010,CanetChateDelamotteWschebor2011}. One has to be careful here however since the KPZ-equation does not as such admit a stationary regime (the interface grows). It is also worth discussing the physical meaning of the symmetry. We emphasize that the stochastic Burgers equation breaks time-reflection symmetry and is truly out-of-equilibrium, but still satisfies a generalized time-reversal symmetry, which is in fact a PT-symmetry. That generalized time-reversal symmetry permits the identification of the fluctuation--dissipation relation we are interested in. This relation was already stated in \cite{ForsterNelsonStephen1977} (see Appendix B there), with the derivation based on the work \cite{DekerHaake1975}, but we could not find in the literature a self-contained derivation of this relation emphasizing the physical aspect in the KPZ context.

\subsection{Setting and Martin-Siggia-Rose action for the stochastic Burgers equation}

Consider the stochastic Burgers equation with a source $\phi(t,x)$ as in Eq.~(\ref{Eq:Main:Burgers}), that we recall here for readability,
\bea  \label{Eq:Burgers2}
&& \partial_t u(t,x) =  \frac{1}{2} \partial_x (u(t,x))^2  + \frac{1}{2} \partial_x^2 u (t,x) -  \partial_x \xi(t,x) + \phi(t,x)   \nn \\
&& u(0,x) = -\mu - \eta(x) \ssp .
\eea
For convenience in this section we take the system on an interval of length $L$ with periodic boundary conditions, i.e., $x \in [-L/2,L/2]$ with $L \in \JR_+$, and we consider the process in a time--window of length $T >0$, i.e., $t \in [0,T]$. We use the short-hands $\int_x := \int_{x \in [-L/2,L/2]}$ and $\int_t := \int_{t \in [0,T]}$. The $L \to \infty$ limit of our results, that is relevant to the application to the continuum directed polymer problem presented in this paper, is thought of as being taken afterwards.
Following the MSR-formalism, and introducing a response field $\tilde{u}(t,x) \in \JR$, the functional probability to observe a field-trajectory $\{ u(t,x) , t \in [0,T] , x \in [-L/2,L/2] \}  $, given the source $\phi$ is formally given by,\footnote{We refer the reader to \cite{AronBiroliCugliandolo2010} for the discussion of the subtleties underlying the derivation of the MSR action, in particular the presence/absence of a Jacobian term in the action.}

\bea
\mathbb{P}_{\mu}^{\phi}[u] && = \langle \delta\left( \partial_t u(t,x) - \frac{1}{2} \partial_x^2 u(t,x) +\partial_x \xi(t,x)  )- \phi(t,x) \right) \rangle   \mathbb{P}^{{\rm ini}}_\mu[u(t=0,.)]  \nn \\
&& = \langle  \int {\cal D}[\tilde{u}]e^{ - \int_{t,x}  i\tilde{u}(t,x) \left( \partial_t u(t,x)  - \frac{1}{2} \partial_x (u(t,x))^2 - \frac{1}{2} \partial_x^2 u(t,x) + \partial_x \xi(t,x)  - \phi(t,x) \right) } \rangle  \mathbb{P}^{{\rm ini}}_\mu[u(t=0,.)] \nn \\
&& = \int {\cal D}[\tilde{u}]e^{ -\int_{t,x}  i \tilde{u}(t,x) \left( \partial_t u(t,x)  - \frac{1}{2} \partial_x (u(t,x))^2 - \frac{1}{2} \partial_x^2 u(t,x) - \phi(t,x)  \right)  - \frac{1}{2} \langle ( \int_{t,x} \tilde{u}(t,x) \partial_x \xi(t,x) )^2 \rangle}  \mathbb{P}^{{\rm ini}}_\mu[u(t=0,.)]  \nn \\
\mathbb{P}_{\mu}^{\phi}[u] && =  \int {\cal D}[\tilde{u}]e^{ -\int_{t,x} i \tilde{u}(t,x) \left( \partial_t u(t,x)   - \frac{1}{2} \partial_x (u(t,x))^2  - \frac{1}{2} \partial_x^2 u(t,x) - \phi(t,x)  \right)  -\frac{1}{2}  \int_{t,x}  (\partial_x \tilde{u}(t,x))^2  }  \mathbb{P}^{{\rm ini}}_\mu[u(t=0,.)]   \ssp .
\eea
And the initial condition is
\bea
\mathbb{P}^{{\rm ini}}_\mu[u(t=0,.)] = Z^{-1} e^{- \frac{1}{2} \int_x ( u(t=0,x) + \mu)^2 }  \ssp ,
\eea
where $Z$ is a $\mu$-independent normalization constant. Hence we have
\bea
\mathbb{P}_{\mu}^{\phi}[u]  =  Z^{-1} \int {\cal D}[\tilde{u}] e^{- S_{\mu}[u,\tilde u] + i \int_{t,x} \tilde{u}(t,x) \phi(t,x) } \ssp ,
\eea
where we have introduced the MSR-action for the theory without source $\phi$ as
\bea \label{Eq:MSRaction}
 && S_\mu[u,\tilde{u}] := S_\mu^{(1)}[u,\tilde{u}]  +S^{(2)}[u,\tilde{u}] +S^{(3)}[u,\tilde{u}]      \\
 && S_\mu^{(1)}[u,\tilde{u}] := \int_{t,x}  i \tilde{u}(t,x)  \partial_t u(t,x)   +\frac{1}{2} \int_x (u(t=0,x) + \mu)^2  \nn \\
 && S^{(2)}[u,\tilde{u}] :=  -  \frac{1}{2} \int_{t,x}  i \tilde{u}(t,x) \partial_x^2 u(t,x)   + \frac{1}{2}  \int_{t,x}  (\partial_x \tilde{u}(t,x))^2 \nn \\
 && S^{(3)}[u,\tilde{u}] := - \frac{1}{2} \int_{t,x} i \tilde{u}(t,x)  \partial_x (u(t,x))^2 \nn \ssp .
\eea
Introducing the average with respect to the MSR-action $\langle \langle \cdots \rangle \rangle_{\mu} := Z^{-1} \int {\cal D}[u , \tilde{u}]  \cdots e^{-S_{\mu}[u, \tilde u]}  $, for any observable of the slope field $O[u]$, we have\footnote{The reason why we introduce a different symbol for the average over the MSR-action is that correlations function involving the response field $\tilde{u}$ do not {\it a priori} have a meaning in the stochastic Burgers theory.}
\bea 
\langle O[u] \rangle_{\mu}^{\phi} \equiv \langle \langle O[u] e^{ i \int_{t,x} \tilde{u}(t,x) \phi(t,x)}  \rangle  \rangle_{\mu} \ssp .
\eea
In particular, introducing as in Section~\ref{Sec:MainResult:LinRes} the response function
\bea
R_{\mu}((t,x) ; (t',x')) :=   \frac{\delta \langle  u(t,x) \rangle_{\mu}^\phi }{ \delta \phi(t',x')} |_{\phi=0} \ssp ,
\eea
we have the formula
\bea \label{Eq:FirstKubo}
R_{\mu}((t,x) ; (t',x')) =  \langle \langle  i \tilde{u}(t',x') u(t,x)   \rangle  \rangle_{\mu} \ssp .
\eea

\subsection{PT and CT symmetries of the MSR action}
We now discuss the time-reversal symmetry of the MSR-action that allows us to relate the response function $R_{\mu}((t,x) ; (t',x'))$ to the two-point correlation function of the field $u(t,x)$ in the absence of a source $\phi$ as in Eq.~(\ref{Eq:MainResultFDT}).

\subsubsection{Qualitative discussion and the transformations at the level of the slope field.}

To make transparent the physical content of the symmetry we adopt here the language of interacting particle systems and think of $\rho(t,x)=1/2+u(t,x)$ as the density of particles in the WASEP (see Section~\ref{Sec:MainResult:SPDES}). We consider three transformations implemented by three operators: time-reversal ${\cal T}$, space-reflection ${\cal P}$ and charge conjugation ${\cal C}$ (or particle-hole transformation). These act on the field $u(t,x)$ as
\be
({\cal T }u)(t,x) = u(T-t,x) \quad , \quad ({\cal P} u)(t,x)  = u(t,-x) \quad , \quad  ({\cal C} u)(t,x)  = -u(t,x) \ssp ,
\ee
and the meaning of these transformation should be clear from the WASEP interpretation. Let us now discuss briefly which of (the composition of) these transformations is a symmetry of the stochastic Burgers equation, i.e., such that $\mathbb{P}_\mu^0[({\cal S }u)] = \mathbb{P}_\mu^0[(u)]$ with ${\cal S}$ a transformation.

\medskip

First, ${\cal T }$ cannot be a symmetry of the stochastic Burgers equation. Indeed, writing (\ref{Eq:Burgers2}) with $\phi \equiv 0$ in the form of a local conservation law
\bea
\partial_t u(t,x) + \partial_x j(t,x) = 0 \quad , \quad j(t,x) = - \frac{1}{2} u(t,x)^2 - \frac{1}{2} \partial_x u(t,x) + \xi(t,x) \ssp ,
\eea
we see that there is a strictly negative current $\langle j(t,x) \rangle_\mu^0 = -\frac{1}{2} \langle u(t,x)^2 \rangle_\mu^0 < 0$. %Indeed, the above continuity equation implies that $j(t,x)$ is odd under time-reversal. If ${\cal T }$ was a symmetry we would thus get that $\langle j(t,x) \rangle_\mu^0 = 0$. 
Hence time-reversal symmetry is forbidden by the presence of the nonlinear term in the stochastic Burgers equation. This is natural since the latter represents the bias in the WASEP. %In the absence of a bias (corresponding to SEP in the language of interacting particle systems) time-reversal symmetry can be restored. % and indeed below we show that ${\cal T}$ can be extended to an operator acting both on the $u$ and $\tilde{u}$ fields of the MSR action and that leaves the first and second part of the action (\ref{Eq:MSRaction}) invariant, but not the third, hence implying time-reversal symmetry for the stochastic Burgers equation in the absence of the non-linear term.
%Thinking again of the interacting particle system, it is clear that there is still room for some generalized time-reversal symmetries to exist. Indeed, 
Reversing time changes the direction of the current of particles. That can however be compensated by a space-reflection. Similarly, exchanging particles and holes also reverses the direction of the current in WASEP. In that case however the mean density is also affected as $\langle \rho \rangle \to 1- \langle \rho \rangle$. In the stochastic Burgers setting, this means that the identity $\mathbb{P}_\mu^0[({\cal P T }u)] = \mathbb{P}_\mu^0[(u)]$ and $\mathbb{P}_\mu^0[({\cal C T }u)] = \mathbb{P}_{-\mu}^0[(u)]$ are to be expected. In what follows, by extending ${\cal C}$ and ${\cal P}$ to operators acting on both fields of the MSR-action we indeed prove the invariance of the action with respect to these transformations from which follows that $\mathbb{P}_\mu^0[({\cal P T }u)] = \mathbb{P}_\mu^0[(u)]$ and $\mathbb{P}_\mu^0[({\cal C T }u)] = \mathbb{P}_{-\mu}^0[(u)]$. These invariance properties will then be used to prove the FDR (\ref{Eq:MainResultFDT}).

\subsubsection{The operators on the MSR-fields and the symmetries}

We now extend the operators ${\cal T}$, ${\cal C}$ and ${\cal P}$ as (writing $(u^{{\cal S}} , \tilde{u}^{\cal S}) = {\cal S} (u , \tilde u)$ for an arbitrary transformation),
\bea
 \left( \begin{array}{c}
  u^{{\cal T}}(t,x) \\
   \tilde{u}^{{\cal T}}(t,x) \end{array} \right) =  \left( \begin{array}{c}
  u(T-t,x) \\
   -  \tilde{u}(T-t,x)  - i ( u(T-t,x) + \mu) \end{array} \right) \ssp , 
\eea
and
\bea
 \left( \begin{array}{c}
  u^{{\cal P}}(t,x) \\
   \tilde{u}^{{\cal P}}(t,x) \end{array} \right) = \left( \begin{array}{c}
  u(t,-x) \\
   \tilde{u}(t,-x)  \end{array} \right) \quad , \quad  \left( \begin{array}{c}
  u^{{\cal C}}(t,x) \\
   \tilde{u}^{{\cal C}}(t,x) \end{array} \right)  =  \left( \begin{array}{c}
  -u(t,x) \\
   -\tilde{u}(t,x)  \end{array} \right) \ssp .
\eea
These different transformation are involutions (${\cal S}^2 =1 $). We have ${\cal P T } ={\cal T P } $, ${\cal C P } ={\cal P C } $ but ${\cal C T_{\mu}} = {\cal C T_{- \mu}} $. Their action on the $u$ fields is basically dictated by the physics of the problem, while their action on the $\tilde{u}$ fields is to satisfy
\bea
& S^{(1)}_{\mu}[u^{\cal T} , \tilde{u}^{\cal T}] = S^{(1)}_{\mu}[u , \tilde{u}] \quad , \quad & S^{(2)}_{\mu}[u^{\cal T} , \tilde{u}^{\cal T}] = S^{(2)}_{\mu}[u , \tilde{u}]    \\
& S^{(1)}_{\mu}[u^{\cal P} , \tilde{u}^{\cal P}] = S^{(1)}_{\mu}[u , \tilde{u}] \quad , \quad & S^{(2)}_{\mu}[u^{\cal P} , \tilde{u}^{\cal P}] = S^{(2)}_{\mu}[u , \tilde{u}]   \\
& S^{(1)}_{\mu}[u^{\cal C} , \tilde{u}^{\cal C}] = S^{(1)}_{-\mu}[u , \tilde{u}] \quad , \quad & S^{(2)}_{\mu}[u^{\cal C} , \tilde{u}^{\cal C}] = S^{(2)}_{\mu}[u , \tilde{u}]  
\eea

These properties hold evidently by remarking that the first and second part of the action can be rewritten as

\bea
 S_\mu^{(1)}[u,\tilde{u}]  = && \int_{t,x}  i ( \tilde{u}(t,x)  + \frac{i}{2} (u(t,x) + \mu) )\partial_t u(t,x)  \nn \\
 && + \frac{1}{4} \left[  \int_x (u(t=0,x) + \mu)^2 +(u(t=T,x) + \mu)^2   \right]  \ssp , \nn \\
 S^{(2)}[u,\tilde{u}]  = &&   - \frac{1}{2}  \int_{t,x} \tilde{u}(t,x) \partial_x^2  ( \tilde{u}(t,x)  + i u(t,x) ) \ssp .
\eea
Physically, in lattice gas language, these are consequences of the fact that the steady symmetric exclusion process is time-reversal invariant, spatially-reflection invariant and that the particle-hole transformation is also a symmetry at half-filling.

We are however here interested in the stochastic Burgers equation, or WASEP, and the third, nonlinear part of the action, breaks the time-reversal symmetry. Indeed,
\bea
S^{(3)}[u^{\cal T} , \tilde{u}^{\cal T}] && = - \frac{1}{2} \int_{t,x} i ( -\tilde{u}(T-t,x) - i(u(T-t,x) + \mu)  \partial_x  (u(T-t,x))^2  \nn \\
&& =  + \frac{1}{2} \int_{t,x} i \tilde{u}(T-t,x) \partial_x  (u(T-t,x))^2 - \int_{t,x} u(T-t,x)\partial_x(u(T-t,x))^2 \nn \\
&& = - S^{(3)}[u , \tilde{u}] \ssp .
\eea
Here the last line follows by changing $t \to T-t$ and remarking that for any periodic function $f(x)$, $\int_x f(x) \partial_x (f(x)^2) = 2\int_x (f(x))^2 \partial_x f(x)  $, but also $\int_x f(x) \partial_x (f(x)^2) =  - \int_x (f(x))^2 \partial_x f(x)  $ (integration by part), and hence $\int_x f(x) \partial_x (f(x)^2) = 0$. The nonlinear part of the action is thus exactly {\it anti-symmetric} under time-reversal. That is a sign of a nonequilibrium steady state with a finite rate of entropy production given by $ \dot{{\rm S}} = \frac{1}{2} \langle\langle  \int_x i \tilde{u}_{t,x} \partial_x (u(t,x))^2 \rangle \rangle$, \cite{MaesRedigMoffaert2000}. The nonlinear part of the action is however also clearly anti-symmetric under charge conjugation (odd number of fields) or parity transformation (odd number of space derivative): $S^{(3)}[u^{\cal P} , \tilde{u}^{\cal P}]=S^{(3)}[u^{\cal C} , \tilde{u}^{\cal C}]= -S^{(3)}[u , \tilde{u}]   $. We retrieve here the fact that reversing time in the ASEP is the same as reversing the sign of the bias. Combining everything we find that
\bea
S_{\mu}[u^{ \cal P \cal T} , \tilde{u}^{ \cal P \cal T}] = S_{\mu} [u , \tilde{u}] \quad \text{and} \quad S_{\mu}[u^{ \cal C \cal T} , \tilde{u}^{ \cal C \cal T}] = S_{-\mu} [u , \tilde{u}]  \ssp .
\eea
Hence ${\cal P T}$ is always a symmetry of the stochastic Burgers equation in the stationary state, while ${\cal C T}$ is only a symmetry in the half-filled case $\mu = 0$ (see next section).

\medskip

{\bf Remark:} At half-filling $\mu=0$, the transformation $({\cal C T} u)(t,x) = -u(-t,x) $ is thus also a symmetry of the stationary Burgers equation. If the field $u(t,x)$ was `chosen' to be odd under time-reversal (hence now call `time-reversal' ${\cal T'} = {\cal C T}$), as would be relevant if $u$ was interpreted as a velocity field, then the stationary Burgers equation would really be invariant under time-reversal symmetry. The action of time-reversal on the field $u$ must however be dictated by the physics, and it is crucial to identify this action correctly since this choice determines if the stationary state should be considered as an out-of-equilibrium one or not. In this section we adopted the WASEP language where it is clear that $u$ should be taken even under time-reversal. Taking $u$ to be even under time-reversal is however also natural in the stochastic growth language where the action of ${\cal T}$ on $h(t,x)$ should be $({\cal T}h)(t,x) = h(T-t,x)$, also leading to a even slope field $u(t,x) = \partial_x h(t,x)$.

\subsection{Stationarity and FDR in the stochastic Burgers equation}

\subsubsection{`Proof' of stationarity}

Let us now use the ${\cal P T}$-symmetry to `prove' the stationarity of the stochastic Burgers equation, a fact which is {\it a priori} not obvious. To this aim we write, using ${\cal S} = {\cal P T}$
\bea \label{Eq:ProofOfStationarity}
\mathbb{P}_\mu^0[u^{\cal S}] && = Z^{-1} \int {\cal D}[\tilde{ u}]  e^{-S_{\mu}[u^{\cal S} ,\tilde{u}]}  \nn \\
&& = Z^{-1} \int {\cal D}[\tilde{ u}^{\cal S}]  e^{-S_{\mu}[u^{\cal S} ,\tilde{u}^{\cal S}]}  \nn \\
&& = Z^{-1} \int {\cal D}[\tilde{ u}]  e^{-S_{\mu}[u ,\tilde{u}]}  \nn \\
\mathbb{P}_\mu^0[u^{\cal PT}] && = \mathbb{P}_\mu^0[u]  \ssp .
\eea
Hence we have the identity $u^{\cal PT}(t,x) = u(T-t,-x) \sim u(t,x) $  in law. In particular $u(T,x) \sim u(0,-x)$ and, using that the initial condition is invariant by space reflection, $u(0,x) \sim u(0,-x)$, we conclude that indeed $u(T,x) \sim u(0,x)$ for all $T$, and the process is stationary. Note that in (\ref{Eq:ProofOfStationarity}) we have crucially used that the change of variables $ u \to \tilde{ u}^{\cal S}$ has a unit Jacobian: ${\cal D}[\tilde{ u}^{\cal S}] = {\cal D}[\tilde{ u}]$. We have also assumed that we could deform the contour of integration on $\tilde{ u}$ back to the real axis after the transformation. Indeed, note that the symmetry ${\cal S}$ does not leave invariant the domain of integration. See \cite{AronBiroliCugliandolo2010} for a discussion of these questions in a similar context.

Note that using now the transformation ${\cal S'} = {\cal C T}$ one similarly obtains $\mathbb{P}_\mu^0[u^{\cal S'}] = \mathbb{P}_{-\mu}^0[u]$: the process with stationary measure $-\mu$ can be generated by applying a time-reflection and a particle-hole transformation to the process with stationary measure $+\mu$.

\subsubsection{`Proof' of the FDR}

Let us now start from the expression of the response function given in (\ref{Eq:FirstKubo}) and use again the ${\cal S} = {\cal P T}$ symmetry:
\bea
R_{\mu}((t,x) ; (t',x')) && =  \langle \langle  i \tilde{u}(t',x') u(t,x)   \rangle  \rangle_{\mu} \nn \\
&&  = \int {\cal D}[u,\tilde{u}] i \tilde{u}(t',x') u(t,x) e^{-S_{\mu}[u,\tilde{u}]} \nn \\
&& = \int {\cal D}[u^{\cal S},\tilde{u}^{\cal S}] i \tilde{u}^{\cal S}(t',x') u^{\cal S}(t,x) e^{-S_{\mu}[u^{\cal S},\tilde{u}^{\cal S}]}  \nn \\
&& = \int {\cal D}[u,\tilde{u}] i \tilde{u}^{\cal S}(t',x') u^{\cal S}(t,x) e^{-S_{\mu}[u,\tilde{u}]} \nn \\
&& =   \langle \langle  i( - \tilde{u}(T-t',-x')  -i ( u(T-t',-x') + \mu)) (u(T-t,-x)) \rangle \rangle_{\mu} \nn \\
R_{\mu}((t,x) ; (t',x')) && =  - R_{\mu}((T-t,-x);(T-t',-x'))   + \langle (u(T-t',-x') + \mu)  u(T-t,-x) \rangle_\mu^0 \nn \ssp .
\eea
Hence, using that the response is causal, i.e., $R_{\mu}((t,x) ; (t',x')) \sim \theta(t-t')$, that there is a space-time translation invariance of the steady state, i.e., $u(t,x) \sim u(t+s,x+y)$ and that $\langle u(t',x')\rangle_{\mu}^{0} = - \mu$, we obtain the FDR
\bea
R_{\mu}((t,x) ; (t',x')) = \theta(t-t')  \left( \langle u(t-t',x-x')  u(0,0) \rangle_{\mu}^0  - \mu^2 \right)\ssp .
\eea
The response is thus also stationary $R_{\mu}((t,x) ; (t',x')) \equiv R_{\mu}(t-t',x-x')$ and the right-hand side is just the two--point connected correlation function of the slope field $u$. Finally, in order to relate that to the correlation function in the theory with $\mu = 0$, note that the change of variables $u(t,x) = - \mu + \hat u(t,x- \mu t)$ maps the stationary stochastic Burgers equation with initial condition $\mu$ (for $u$) to the stationary stochastic Burgers equation with initial condition $\mu=0$ (for $\hat u$). Hence 
\bea
 \langle u(t-t',x-x')  u(0,0) \rangle_\mu^0  && =  \langle (-\mu+ u(t-t',x-x' - \mu(t-t'))) ( -\mu  +u(0,0) )  \rangle_0^0 \nn \\ 
 &&  = \mu^2 + \langle u(t-t',x-x' - \mu(t-t')) u(0,0)  \rangle_0^0
\eea
We thus obtain that
\bea
R_{\mu}(t-t' , x-x')  =  R_0 (t-t' , x-x'-\mu(t-t'))
\eea
with
\bea
R_0 (t-t' , x-x') = \theta(t-t') \langle u(t-t',x-x')  u(0,0) \rangle_{0}^0  \ssp .
\eea
which is the result presented in Eq.~(\ref{Eq:MainResultFDT}).

\section{The directed polymer midpoint distribution } \label{Sec:ApplicationToDP}

\subsection{Connecting the midpoint distribution with Burgers linear response function} \label{Sec:ApplicationToDP:MainRes}

We now consider the directed polymer partition sum $Z(t,x)$ with stationary initial condition and in the presence of a small perturbing potential as presented in Section~\ref{Sec:MainResult:ModelAndUnits}. Using the path-integral representation (\ref{Eq:DefZPI}) it is clear that the quenched pdf $q_{\mu}((t',x')|(t,x))$ that the DP passes through $(t',x')$, given that it passes through $(t,x)$ introduced in Eq.~(\ref{Eq:Defq}) is just given by, in a fixed random environment,
\bea
q_{\mu}((t',x')|(t,x)) = -\frac{\delta \ln Z(t,x)}{\delta V(t',x')}|_{V=0}  \ssp .
\eea
Its average over disorder, $p_{\mu}((t',x')|(t,x))$, is thus a response function,
\bea\label{iad}
p_{\mu}((t',x')|(t,x)) = \tilde{R}_\mu((t,x);(t',x')):= -\frac{ \langle \delta \ln Z(t,x) \rangle_\mu^V }{\delta V(t',x')}|_{V=0}  \ssp .
\eea
We next show that this response function is the same as the one introduced in the stochastic Burgers setting, see Eq.~(\ref{Eq:Main:DefR}). To that aim note that using $u(t,x) = \partial_x \ln Z(t,x)$ we can write, for any $y \in \JR$
\bea
\tilde{R}_\mu((t,x);(t',x'))-\tilde{R}_\mu((t,y);(t',x')) = - \int_y^x dz \frac{\delta \langle  u(t,z) \rangle_{\mu}^V}{\delta V(t',x')}|_{V=0} \ssp .
\eea
On the other hand $R_{\mu}(t-t',x-x')$ gives the linear response of $\langle u(t,x) \rangle_\mu^\phi$ for a small source $\phi(t',x') = -\partial_{x'} V(t',x')$. This means that to first order in $V$,
\bea
\langle  u(t,x) \rangle_{\mu}^V-\langle  u(t,x) \rangle_{\mu}^0 && = -  \int_{0}^t dt' \int_{x'} dx' R_\mu(t-t',x-x' )  \partial_{x'} V(t',x')  +O(V^2) \nn \\
&& =  - \int_{0}^t dt' \int_{x'} dx' \partial_{x} R_\mu(t-t' , x-x') V(t',x')  + O(V^2) \ssp .
\eea
Hence $\frac{\delta \langle  u(t,z) \rangle_{\mu}^V}{\delta V(t',x')}|_{V=0}  = - \partial_z R_\mu(t-t',z-x')$ and
\bea
\tilde{R}_\mu((t,x);(t',x'))-\tilde{R}_\mu((t,y);(t',x'))  && = + \int_y^x dz \partial_z R_\mu(t-t',z-x') \nn \\
&& = R_\mu(t-t',x-x')-R_\mu(t-t',y-x')   \ssp .
\eea
Sending $y \to \infty$ with the other arguments fixed and using that $\lim_{y \to \infty} \tilde{R}_\mu((t,y);(t',x'))  = \lim_{y \to \infty} R_\mu(t-t',y-x') = 0  $, we obtain
\bea
\tilde{R}_\mu((t,x);(t',x')) = R_\mu(t-t',x-x') \ssp .
\eea
Thus Eq.~(\ref{Eq:MainLinResDP}) follows and, combining that with \eqref{iad} we obtain $p_\mu((t',x')|(t,x)) = R_{\mu}(t-t',x-x')$ and the results presented in Section~\ref{Sec:MainResult:Midpoint1}.

\medskip

{\bf Remark} At the physics level of rigor, this provides a `proof' that the response function $R_{\mu}(t-t',x-x')$ is in fact a pdf. The fact that it is normalized to unity, i.e., that $\int_{x'} dx' R_{\mu}(t-t',x-x') = \int_{x} dx R_{\mu}(t-t',x-x')  =1$ is also natural from its definition as a response function. Indeed, considering again the stochastic Burgers equation perturbed by a source $\phi$ (\ref{Eq:Main:Burgers}) with, for simplicity, periodic boundary conditions on a ring of length $L$, one easily obtains that $N(t) = \int_{-L/2}^{L/2} u(t,x) dx$ ($\sim$ total number of particles in the WASEP interpretation) satisfies
\bea
\partial_t N(t) = \int_0^L \phi(t,x)  \Longrightarrow N(t)-N(0) = \int_0^t dt' \int_0^L dx  \phi(t',x)  \ssp .
\eea
As a consequence,
\bea
\int_{-L/2}^{L/2} dx R_{\mu}(t-t',x-x') = \frac{\delta \langle N(t) \rangle_\mu^\phi}{\delta \phi(t',x')} |_{\phi=0} = \theta(t-t') \ssp .
\eea
Hence the normalization of $R_{\mu}(t-t',x-x')$ has here its origin in the fact that $u(t,x)$ is a conserved field if the perturbation $\phi(t,x)$ is not present. Seeing that $R_{\mu}$ is positive from its definition as a response function is obvious without the nonlinearity in the dynamics, but it also seems intuitively reasonable for the nonlinear case.

\subsection{Large time limit and interpretation in terms of the Airy process} \label{Sec:ApplicationToDP:Airy}

Let us now make the connection between our result and the Airy process ${{\cal A}}_2(x)$ as summarized by Eq.~(\ref{Eq:MainRes:Airy}). For that purpose let us first introduce the partition sum of the continuum directed polymer with no external potential $V\equiv 0$, the initial point at $(t_i,x_i)$ and the final point at $(t_f,x_f)$ with $t_f>t_i$. By definition, admissible polymer paths are thus continuous functions $\pi : [t_i,t_f] \to \JR$ with $\pi(t_i) = x_i$ and $\pi(t_f)=x_f$ and the point-to-point partition sum of this directed polymer is
\bea \label{Eq:DefZptop}
Z_{t_i,x_i}^{t_f,x_f} := \int_{\pi(t_i) = x_i}^{\pi(t_f)  = x_f}  {\cal D}[\pi] e^{- \int_{0}^t ds \left( \frac{1}{2} (\partial_s \pi(s))^2  + \xi(s,\pi(s))  \right)   }  \ssp .
\eea
In the remaining of this section we take $t_i = x_f = 0$,  $t_f=t$ and $x_i = x \in \JR$. Reintroducing a two-sided Brownian motion without drift $B_0(x) =\int_x \eta(x)$ living on the first time, we can now rewrite the pdf $p_0(t,x)$ as the average over disorder
\bea
p_0(t,x) =  \left\langle  \frac{e^{-B_0(x)} Z_{0,x}^{t,0}  }{ \int_{x' \in \JR} dx' e^{-B_0(x')} Z_{0,x'}^{t,0}  } \right\rangle_0^0 \ssp .
\eea
The Airy-Process now intervenes through the assumption that the following equality in law holds at large time in the KPZ scaling limit $x \sim t^{2/3}$ (see e.g. \cite{QuastelRemenik2014}),
\bea
Z_{0,x}^{t,0}\sim \frac{1}{\sqrt{2 \pi t}} e^{-\frac{x^2}{2 t} - \frac{t}{24} + 2^{-1/3} t^{1/3} {{\cal A}}_2(2^{-1/3} t^{-2/3} x )} 
\eea
Rescaling as before $x \to (2 t^2)^{1/3}$ and rescaling the two-sided Brownian motion as $ B_0((2 t^2)^{1/3} y) = (\sqrt{2} t)^{1/3} B(y) $ with $B(y)$ a unit double-sided Brownian motion, we see that
\bea
{\cal P}(t,y) && = (2 t^2)^{1/3} p_0(t , (2 t^2)^{1/3} y) \simeq \langle   \frac{e^{2^{-1/3} t^{\frac{1}{3}} \left(   {\cal A}_2(y)  - y^2 - \sqrt{2} B(y) \right) }}{ \int_{z \in \JR} dz e^{2^{-1/3} t^{\frac{1}{3}} \left(   {\cal A}_2(z)  - z^2 - \sqrt{2} B(z)) \right) } } \rangle_0^0 \ssp .
\eea
In the infinite time limit $ t\to \infty$ the integrals are dominated by their optimum and we thus obtain
\bea
{\cal P}_\infty(y) = \langle \delta\left(y  -  {\rm argmax}_{z \in \JR} \left[   {\cal A}_2(z)  - z^2   -  \sqrt{2} B(z) \right] \right) \rangle_0^0 \ssp ,
\eea
which is equivalent to Eq.~(\ref{Eq:MainRes:Airy}).

\subsection{Midpoint distribution of the DP in the stationary regime} \label{Sec:ApplicationToDP:PointToPoint}
We now argue that Eq.~(\ref{Eq:Main:BeyondStat}) holds. Starting from the definition (\ref{Eq:Main:BeyondStat00}), we decompose the DP paths in a part from $t_i \to t'$ and a part from $t' \to t$. We obtain
\bea \label{Eq:MidpointInStat1}
p((t',x') |(t_i,x_i) , (t,x)) = \langle \frac{ Z_{t_i,x_i}^{t',x'}  Z_{t',x'}^{t,x}  }{\int_{y \in \JR} dy Z_{t_i,x_i}^{t',y}  Z_{t',y}^{t,x} } \rangle = \langle \frac{ Z_{t',x'}^{t,x}  }{\int_{y \in \JR} dy \frac{Z_{t_i,x_i}^{t',y}}{Z_{t_i,x_i}^{t',x'}}  Z_{t',y}^{t,x} } \rangle   \ssp .
\eea
where we have used again the definition of the point-to-point DP-partition sum (\ref{Eq:DefZptop}). Taking now $t_i \to -\infty$ with $x_i = \varphi t_i$, one expects that the KPZ-height variable $ \ln\left( \frac{Z_{t_{}i,x_i}^{t',y}}{Z_{t_i,x_i}^{t',x'}} \right) $ locally converges to one of its stationary state. I.e., we expect
\bea
\ln\left( \frac{Z_{t_i,\varphi t_i}^{t',y}}{Z_{t_i,\varphi t_i}^{t',x'}} \right) \sim_{t_i \to -\infty}  -B_{\mu}(y) +B_{\mu}(x') \ssp ,
\eea
with $B_\mu(x) = \mu x + \int_{0}^x dx' \eta(x')$ a double-sided Brownian motion as in the definition of the model with stationary initial condition. It remains to identify the correct parameter $\mu$. That is done by averaging over the disorder and by using that 
\bea
\langle -B_{\mu}(y) + B_{\mu}(x') \rangle = \mu(x'-y) && = \lim_{t_i \to -\infty} \left( \langle  \ln\left( Z_{t_i,\varphi t_i}^{t',y}\right)   -   \ln\left( Z_{t_i,\varphi t_i}^{t',x'}\right)  \rangle  \right) \nn \\
&& = \lim_{t_i \to -\infty} \left( - \frac{(y-\varphi t_i)^2}{(t'-t_i)}  +  \frac{(x'-\varphi t_i)^2}{(t'-t_i)} \right) \nn \\
&& = \varphi ( x'-y) \ssp ,
\eea
where the right-hand side of this calculation uses the well-known statistical-tilt symmetry of the continuum DP-model that states the identity in law $Z_{t_i ,x_i}^{t_f, x_f} \sim e^{- \frac{(x_f-x_i)^2}{2(t_f-t_i)}} Z_{0,0}^{t_f-t_i,0} $ (see e.g. \cite{LeDoussalCalabrese2012}).

\medskip

We therefore get the correct parameter as $\mu(\varphi) := +\varphi$. Substituting into (\ref{Eq:MidpointInStat1}) we obtain
\bea
\lim_{t_i \to - \infty} p((t',x') |(t_i,\varphi t_i) , (t,x)) =   \langle   \frac{ Z_{t',x'}^{t,x}  }{\int_{y \in \JR} dy e^{-B_{\mu = \varphi}(y)} Z_{t',y}^{t,x} } \rangle   \ssp .
\eea
And the right-hand side is indeed equal to $p_{\mu=\varphi}((t',x') |(t,x))$, which gives Eq.~(\ref{Eq:Main:BeyondStat}).

\subsection{Results on the Log-Gamma polymer} \label{Sec:ApplicationToDP:LG}

Here we give some details on the derivation of the results presented in Section~\ref{Sec:Main:LG}.

\medskip

{\bf Weak universality} Following Section~III.2.2.b of \cite{ThieryPHD} we know that the weak-noise limit of the Log-Gamma polymer converges to the continuum DP as follows: taking
\bea
\st = \gamma^2 t \quad , \quad \sx = \frac{\gamma}{2} x \quad , \quad \gamma \to \infty \ssp ,
\eea
we have the convergence in law
\bea
\lim_{\gamma \to \infty} e^{\gamma^2 t \ln(\gamma/2) } \sZ(\st= \gamma^2 t , \sx = \frac{\gamma}{2} x )  = Z(t,x) \ssp .
\eea
Adapting this rescaling to the probability $\spp(\st,\sx)$ this leads to (\ref{Eq:MainRes:WeakU}). 

\smallskip

{\bf Strong universality } As mentioned in Section~\ref{Sec:Main:LG}, the parameters $\kappa_{\sx} , \kappa_{\st}>0$ should be chosen in such a way that: (i) the curvature of the free energy of the point-to-point discrete and continuum model coincide in the scaling limit; (ii) the variance of the differences of free energies between two different points in the stationary models coincide.

\smallskip

Let us start with (ii). In the Log-Gamma polymer, the stationarity implies that at each $\st\geq 0$, $\ln \sZ(\st,\sx+1) - \ln \sZ(\st,\sx) \sim \ln(u)  -\ln(v) $ where $u$ and $v$ are iid RVs distributed as $u\sim v \sim (Gamma(\gamma/2))^{-1}$. Hence,
\bea \label{Eq:AppliLG1}
\langle (\ln \sZ(\st,\sx) - \ln \sZ(\st,0))^2 \rangle  = 2 \sx \psi'(\gamma/2) \ssp , 
\eea
where here we have used that if $u \sim (Gamma(\gamma/2))^{-1}$, $\langle u^2 \rangle-\langle u \rangle^2 = \psi'(\gamma/2)$. In the continuum model on the other hand we have from stationarity 
\bea \label{Eq:AppliLG2}
\langle (\ln Z(t,x) - \ln Z(t,0))^2 \rangle = \langle (B_0(x) - B_0(0))^2 \rangle = x \ssp .
\eea
Therefore, identifying (\ref{Eq:AppliLG1}) and (\ref{Eq:AppliLG2}) with $\sx = \kappa_{\sx} x$ requires to take $\kappa_{\sx} = \frac{1}{2\psi'(\gamma/2) } $ as in Eq.~(\ref{Eq:Main:RescalingLogGamma2}).

\smallskip

For (i), we now need to consider the point-to-point model. In the continuum case, the point-to-point partition sum $Z_{0,0}^{t,x}$ was already defined in Eq.~(\ref{Eq:DefZptop}) and we have already recalled in Sec.~\ref{Sec:ApplicationToDP:PointToPoint} that the statistical-tilt-symmetry guarantees that
\bea \label{Eq:AppliLG3}
\langle \ln Z_{0,0}^{t,x} -\ln Z_{0,0}^{t,x} \rangle = -\frac{x^2}{2t} \ssp ,
\eea
which is an exact result $\forall (x,t)$. In the discrete model, we introduce in the same way the point-to-point partition sum $\sZ_{0,0}^{\st,\sx}$ as, taking back the notation of Section~\ref{Sec:Main:LG}:
\bea 
\sZ_{0,0}^{\st,\sx} := \sum_{\pi, \pi(0) = 0 , \pi(\st)=\sx} \prod_{\sss = 0}^\st w_{\sss}(\pi(\sss))  \ssp .
\eea
In this model it is known that for large $\st$ and $\sx = o(\st)$ (relevant to the scaling limit $\sx \sim \st^{2/3}$), we have (see Eq.~(79) in \cite{ThieryLeDoussal2014})
\bea \label{Eq:AppliLG4}
\langle \ln \sZ_{0,0}^{\st,\sx} -\ln \sZ_{0,0}^{\st,\sx} \rangle \sim \frac{\sx^2}{\st} \frac{2 \psi'(\gamma/2)^2}{\psi''(\gamma/2)^2} \ssp .
\eea
Hence, matching now (\ref{Eq:AppliLG3}) and (\ref{Eq:AppliLG4}) with $\sx = \kappa_\sx x$ and $\st  = \kappa_\st t$ requires to take $\kappa_{\st} = -\frac{4 \psi'(\gamma/2)^2}{\psi''(\gamma/2)} \kappa_\sx^2$, which leads to (\ref{Eq:Main:RescalingLogGamma2}).

\medskip

\begin{figure}
\centerline{\includegraphics[width=6cm]{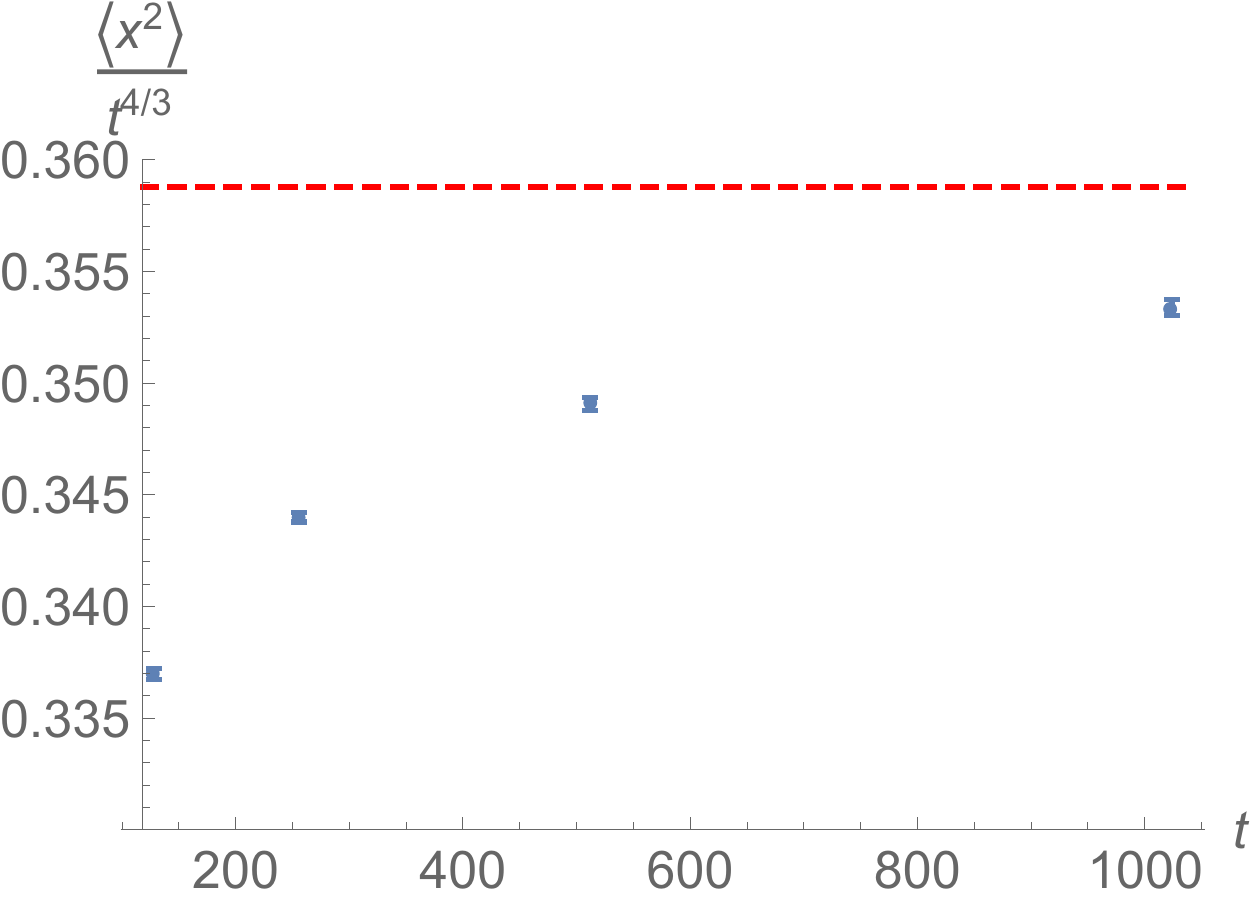} \quad  \includegraphics[width=6cm]{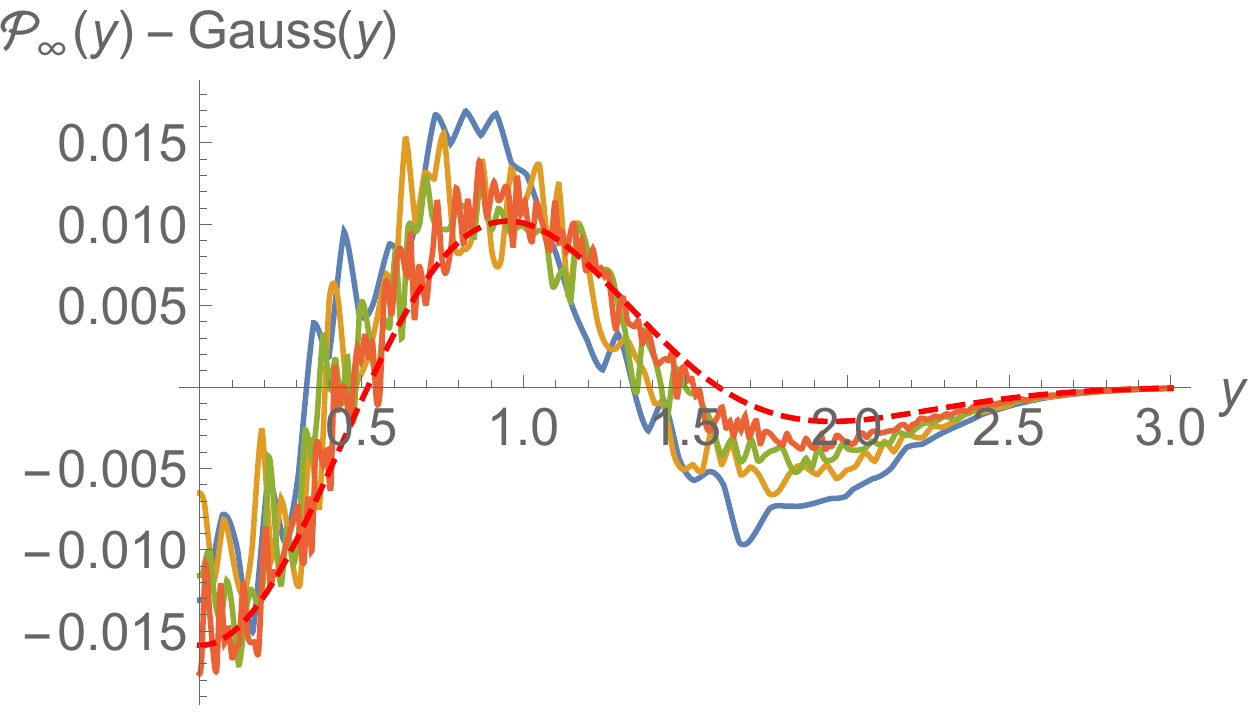}} 
\caption{Left: The blue dots give the numerical evaluation of the rescaled second moment of the midpoint distribution $\spp_0(\st,\sx)$ in the Log-Gamma polymer for polymers of length $\st= 128,256,512,1024$. The red-dashed line is the asymptotic prediction that uses (\ref{Eq:MainRes:StrongU}) together with the numerical evaluation of the variance of $f_{{\rm KPZ}}(y)$ as in Section~\ref{Sec:MainResult:Midpoint2}. Error-bars are $3-\sigma$ estimates. Right: Plain blue, yellow, green and orange lines: evaluation of the rescaled midpoint distribution $\spp_0(\st,\sx)$ (rescaled as in (\ref{Eq:MainRes:StrongU})) minus $Gauss(y)$, a Gaussian distribution with the same variance as $f_{{\rm KPZ}}(y)$. Red-dashed line: evaluation of $f_{{\rm KPZ}}(y)-Gauss(y)$.}
\label{fig:Simu2}
\end{figure}

{\bf Details on the simulations and supplementary results} The simulations were performed using the JULIA language and a standard transfer matrix algorithm to compute the different partition sums involved in the definition of the midpoint probability in the stationary Log-Gamma polymer $\spp_0(\st,\sx)$, cf Eq.~(\ref{Eq:Main:DefpLG}). We performed simulations for polymers of length $\st \in \{ 128,256,512,1024\}$. Averages over disorder were approximated by sampling $10^7$ independent realization of the disorder using the `Distributions' package of JULIA. Besides the result presented in Section~\ref{Sec:ApplicationToDP:LG}, we also present here: (i) On the left of Fig.~\ref{fig:Simu2} a comparison between the numerical evaluation of the second moment of $\spp_0(\st,\sx)$ and the asymptotic prediction (\ref{Eq:MainRes:StrongU}); (ii) On the right of Fig.~\ref{fig:Simu2} an illustration of the convergence of (the rescaled version of) $\spp_0(\st,\sx)$ to its asymptotic form given by the prediction (\ref{Eq:MainRes:StrongU}), further emphasizing the difference between the prediction and a simpler Gaussian Ansatz: (iii) A numerical evaluation of the tail of the distribution ${\cal P}_{\infty}(y)$ using again the results (\ref{Eq:MainRes:StrongU}), and a comparison with the prediction and a simpler Gaussian Ansatz.

\begin{figure}
\centerline{ \includegraphics[width=6cm]{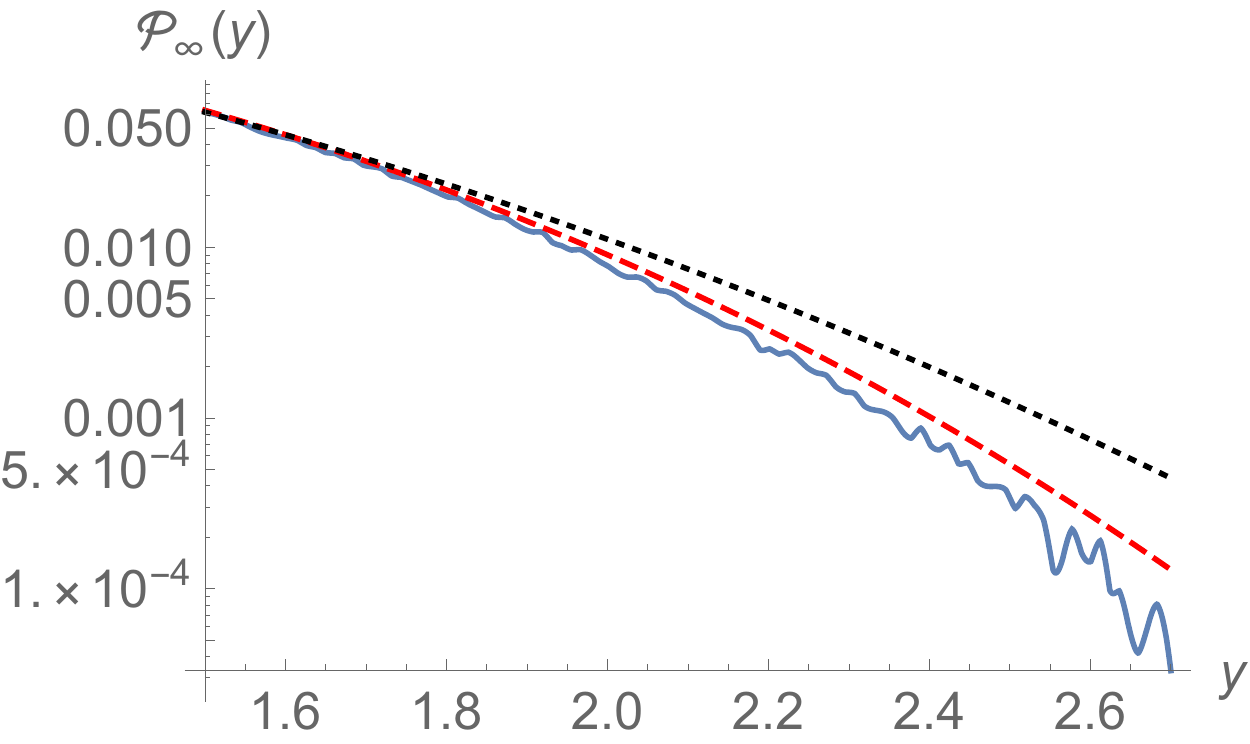}} 
\caption{Logarithmic plot of (i) Blue line: the numerical approximation of ${\cal P}_\infty(y)$ using the result (\ref{Eq:MainRes:StrongU}) and the simulations of the Log-Gamma polymer for polymers of length $\st = 1024$; (ii) Red-dashed line: the $f_{{\rm KPZ}}(y)$ function; (iii) Black-dotted line: a Gaussian distribution with the same variance as the $f_{{\rm KPZ}}(y)$ function.}
\label{fig:Simu3}
\end{figure}

\section{Summary and outlook}

The midpoint pdf of the continuum directed polymer in the stationary regime, when averaged over the disorder, was shown to be given by the two space-time points correlation function of the field in the stationary stochastic Burgers equation.  Complemented with the exact result of Imamura-Sasamoto in \cite{ImamuraSasamoto2013}, this provides an explicit formula for that distribution. In the KPZ scaling limit this pdf converges to the $f_{{\rm KPZ}}(y)$ scaling function introduced by Pr\"ahofer and Spohn in \cite{PrahoferSpohn2004}. Interpreting this result in the framework of KPZ universality, this provides a characterization of the Pr\"ahofer and Spohn scaling function in terms of a natural variational problem involving the Airy process. Still using KPZ universality, we obtained a direct numerical check of this result using simulations of the Log-Gamma polymer.

\medskip

  The main observation that makes all that possible is (i) the understanding of the midpoint distribution as a response function  in the stochastic Burgers equation, and (ii) the proof of a fluctuation--dissipation relation for the stochastic Burgers equation.

 \medskip

Let us mention some further directions of research. First, it is interesting to understand if similar useful fluctuation--dissipation relations can be established for other models in the KPZ universality class. In particular some discrete exactly solvable models of directed polymer on the square lattice have a stationary measure (in the same weak sense that in the continuum) with a generalized time--reversal invariance property similar to the continuum one \cite{Seppalainen2009,Thiery2016}. That would open the way to a rigorous proof of the results of this paper. Another research direction is to try to extend our calculations to the continuum directed polymer beyond the stationary regime. The relation between the midpoint distribution of the directed polymer and a linear response function remains valid. Yet, the calculation of this response function in terms of correlations in the unperturbed theory will be more difficult since it will also involve an additional `frenetic' term as found in the theory of linear response out-of-equilibrium \cite{BaiesiMaesWynants2009,BaiesiMaes2013}.

\begin{acknowledgements}

T.T. is grateful to Vivien Lecomte and Pierre Le Doussal for stimulating discussions. T.T. has been supported by the InterUniversity Attraction Pole phase VII/18 dynamics, geometry and statistical physics of the Belgian Science Policy.

\end{acknowledgements}

\appendix
\section{Imamura-Sasamoto result}\label{app:IS}

In this appendix we recall for completeness the result obtained by Imamura and Sasamoto in \cite{ImamuraSasamoto2013} for the analytical expression of the Burgers stationary two-point correlation function. We note that another (presumably equivalent) result could be obtained by using the formulas  in the mathematically rigorous work \cite{BorodinCorwinFerrariVeto2015}. In \cite{ImamuraSasamoto2013} Imamura and Sasamoto consider the KPZ equation with the convention
\bea
\partial_t h(t,x) = \frac{\lambda}{2} (\partial_x h(t,x))^2 + \nu \partial_x^2 h(t,x)+ \sqrt{D}\xi(t,x) \ssp ,
\eea 
and we will thus take $D=2 \nu =\lambda = 1$ in their result to conform with our conventions; see Eq.~(\ref{Eq:Main:KPZ}). Their result is given in terms of the scaling function $g_t(y)$ which is defined as\footnote{Here we note that there seems to be some misprints in the arXiv version (v1) of \cite{ImamuraSasamoto2013}, and here we follow the published version.}

\bea
g_t(y) = \int_{-\infty}^\infty s^2 \frac{d F_{w=0,t}(s;y)}{ds} ds- \left(  \int_{-\infty}^\infty s \frac{d F_{w=0,t}(s;y)}{ds} ds\right)^2 \ssp .
\eea
And for each $(t,y)$, $F_{w=0,t}(s;y)$ is the cumulative distribution function of the fluctuations of the KPZ interface at $(t,y)$: $g_t(y)$ is the variance of the height at $(t,y)$. The expression for $F_{w=0,t}$ is, using their parameters $\alpha$ and $\gamma_t$ are $\alpha=1$ and $\gamma_t=(t/2)^{1/3})$ (see Theorem 2 in \cite{ImamuraSasamoto2013})
\bea \label{Eq:App:IM1}
F_{w=0,t}(s;X) = \frac{d/ds}{\Gamma(1+\gamma_t^{-1} d/ds)} \int_{\JR} du e^{-\gamma_t(s-u)} \left(\nu_{w=0,t}(u;X) -\nu^{(\delta)}_{w=0,t}(u;X)\right) \ssp ,
\eea
with
\bea
&& \nu_{w=0,t}(u;X) = {\rm Det}(I - A_{-X,X})L_{-X,X}(u) + {\rm Det}(I - A_{-X,X} - D_{-X,X})  \nn \\
&& \nu_{w=0,t}^{(\delta)}(u;X) = {\rm Det}(I - A_{-X,X}^{(\delta)})L_{-X,X}^{(\delta)}(u) + {\rm Det}(I - A_{-X,X}^{(\delta)} - D_{-X,X}^{(\delta)}) \ssp , 
\eea
and
\bea
&& A_{-X,X}(\xi_1 , \xi_2) = C_t(\xi_1) \int_{u}^{\infty} dy Ai_{\Gamma}^{\Gamma}(\xi_1 + y ,\frac{1}{\gamma_t} , 1- \frac{X}{\gamma_t} , 1+ \frac{X}{\gamma_t}) Ai_{\Gamma}^{\Gamma}(\xi_2 + y ,\frac{1}{\gamma_t} , 1+ \frac{X}{\gamma_t} , 1-\frac{X}{\gamma_t}) \nn \\ 
&& D_{-X,X}(\xi_1,\xi_2) = (A_{-X,X}C_t B_{-X,X,u})(\xi_1)B_{X,-X,u}(\xi_2) \nn \\
&& L_{X,-X}(u) = - \frac{2\gamma_E}{\gamma_t} + u -X^2 -1 \nn \\
&& +  \int_{\JR} dx C_t(x)\left(B_{-X,X,u}^{(1)}(x) B_{X,-X,u}^{(2)}(x) + B_{X,-X,u}^{(1)}(x) B_{-X,X,u}^{(2)}(x)  - B_{-X,X,u}^{(2)}(x) B_{X,-X,u}^{(2)}(x) \right)  \nn \\
&& B_{-X,X,u}^{(1)}(x) = e^{-X^3/2 + (x+u) X} \quad , \quad B_{-X,X,u}^{(2)}(x) = \int_{0}^{\infty} d\lambda e^{-X\lambda} Ai_{\Gamma}^{\Gamma}(x + u + \lambda , \frac{1}{\gamma_t} , 1 + \frac{X}{\gamma_t} , 1  - \frac{X}{\gamma_t})  \nn \\
&& B_{-X,X,u}(x) = B_{-X,X,u}^{(1)}(x) - B_{-X,X,u}^{(2)}(x) \quad, \quad C_t(x) = \frac{e^{\gamma_t x }}{e^{\gamma_t x}-1} \ssp .
\eea
The expression with the superscript $\delta$ are identical with $C_t(x) \to C_t(x)^{(\delta)} = C_t(x) - \delta(x)$. $\gamma_E$ in the expression of $L_{X,-X}(u)$ denotes Euler's gamma constant. Finally the function $Ai_{\Gamma}^{\Gamma}$ is a deformed Gamma function defined by
\bea
Ai_{\Gamma}^{\Gamma}(a,b,c,d) = \frac{1}{2\pi} \int_{\Gamma_{id/b}} dz e^{i z a + i \frac{z^3}{3}} \frac{\Gamma(ib z + d)}{\Gamma(-ibz + c)}
\eea 
with $\Gamma_{id/b}$ a contour from $-\infty$ to $+\infty$ passing below $i d/b$. In (\ref{Eq:App:IM1}) $\frac{1}{\Gamma(1 + \gamma_t^{-1} d/ds)}$ is the operator defined by the Taylor expansion of the Gamma function around $1$, and ${\rm Det}(I-K) $ denotes the Fredholm determinant of the operator with the kernels defined above, and $(\xi_1,\xi_2)\in \JR^2$. No matter how complicated this result seems it can indeed be plotted, see Fig.~3 in \cite{ImamuraSasamoto2013}. In the limit $t \to \infty$ the results simplifies a bit. In that case $g_{\infty}(y)$ is still given by a variance, but now associated with the cumulative distribution function $F_{w=0}(s;y)=\lim_{t\to\infty}F_{x=0,t}(s;y)$. An explicit expression for $F_{w=0}(s;y)$ was given in Theorem~3 of \cite{ImamuraSasamoto2013}.

\medskip

Let us finally make the connection with our setting. In \cite{ImamuraSasamoto2013} $g_t(y)$ is related to Burgers' stationary two-point correlation through (see around Corollary~4)
\bea
g_t(y)=  \left( \frac{2}{ t} \right)^{2/3} C(t , (2 t^2)^{1/3} y) \ssp ,
\eea
and
\bea
\langle u(t,x) u(0,0) \rangle_0^0 = \frac{1}{2} \partial_x^2 C(t,x) \ssp .
\eea
Following our result (\ref{Eq:MainResultFDT}) and the rescaling (\ref{Eq:MainResultRRescaling}) one thus easily arrives at Eq.~(\ref{Eq:MainResultFDTRescaled}).

\bibliographystyle{hieeetr}

\bibliography{Citations-Thimothee-06.10.16.bib}
\end{document}